\documentclass[journal]{IEEEtran}

\usepackage[utf8]{inputenc}
\usepackage[T1]{fontenc}

\usepackage{graphics} 
\usepackage{epsfig} 
\usepackage{amsmath} 
\usepackage{amssymb}  
\usepackage[sorting=none,style=ieee]{biblatex}
\usepackage[caption=false,font=footnotesize]{subfig}
\usepackage{xcolor}
\usepackage{booktabs}

\title{Modulating Retroreflectors for CubeSat Optical Inter Satellite Links: Modeling, Optimization, and Benchmarking}

\author{Makafui Avevor,
Hossein Safi, \emph{Member}, \emph{IEEE},\\
Harald Haas, \emph{Fellow}, \emph{IEEE}, and
Iman Tavakkolnia, \emph{Senior Member}, \emph{IEEE}
\thanks{H. Safi, H. Haas, and I. Tavakkolnia are with the LiFi R\&D Center (LRDC), the University of Cambridge, CB3 0FA, Cambridge, UK, (e-mails: \{hs905, huh21, it360\}@cam.ac.uk). M. Avevor is with Information and Network Science Lab, the University of Oxford, Oxford, UK, (e-mail: makafui.avevor@linacre.ox.ac.uk).
This work was supported by the Future Telecoms Research
Hubs, Platform for Driving Ultimate Connectivity (TITAN), sponsored by the Engineering and Physical Sciences Research Council (EPSRC) under Grant EP/X04047X/1 and EP/Y037243/1. Corresponding author: Iman Tavakkolnia}}

\begin{document}

\maketitle
\thispagestyle{empty}
\pagestyle{empty}

\begin{abstract}
Modulating retroreflectors (MRRs) offer a promising pathway to low-complexity and energy efficient asymmetric optical inter-satellite link (OISL) for small spacecrafts, such as CubeSats. In this paper, we develop a unified statistical channel model for an on–off keying modulated, retroreflector-enabled OISL. The model captures both stochastic and deterministic pointing losses, as well as signal-dependent noise. Stochastic channel distributions are approximated via Monte Carlo simulation, and system optimization is carried out under CubeSat constraints using the achievable information rate as the primary metric. In addition, we derive bit-error ratio and outage probability to evaluate communication reliability. The proposed architecture is benchmarked against three state-of-the-art CubeSat laser terminals, i.e., NASA’s Optical Communications and Sensors Demonstration (OCSD), DLR’s OSIRIS4CubeSat, and NASA’s CLICK B/C. Results indicate that an optimized MRR-based transmitter can outperform OCSD and achieve performance comparable to OSIRIS4CubeSat at ranges below 500 km, while consuming only 2.5 W of power during transmission, significantly less than conventional CubeSat optical terminals.
\end{abstract}


\section{INTRODUCTION}

CubeSat technology has emerged as an enabler of modern space research and technology demonstration \cite{saeed2020cubesat}. Their standardized form factors, low development costs, and short development cycles make them well suited for rapid prototyping of space technologies, such as advanced communication systems in low Earth orbit (LEO) \cite{CubeSat}. Increasingly sophisticated CubeSat missions now carry high-resolution sensors and scientific instruments that generate large volumes of data. However, with typical lifetimes of less than five years at altitudes below 400 km, these missions place exceptional demands on the communication subsystem to maximize return within limited operational windows. As a result, high-throughput reliable communication has become a crucial requirement for CubeSat missions \cite{al2024improving}.

Conventional CubeSats rely primarily on radio frequency (RF) links for communications. Their maturity and relaxed pointing requirements have made RF the defualt standard. However, most CubeSat downlinks still operate in the Amateur Radio bands, constraining data rates to only a few kilobits per second. While S- and X-band systems can achieve up to 10–100 Mbps \cite{Ka_CubeSat}, cost and spectrum scarcity, particularly in near-Earth operations, remains a limiting factor. Even with recent advances in attitude determination and control systems (ADCS) enabling Ka-band operation with sub-degree pointing accuracy, throughput typically remains in the hundreds of Mbps \cite{Ka_pointing}. These capabilities fall short of the gigabit-class data rates increasingly demanded by missions.

Laser-based optical communications offer a compelling alternative in this domain. By trading the relaxed pointing tolerance of RF systems for significantly higher gain, optical links enable device miniaturization, longer ranges, and gigabit-per-second data rates, all within CubeSat size, weight, and power (SWaP) constraints \cite{RF_pointing}. The optical spectrum is unlicensed and large. However, the main barriers to CubeSat-scale optical communications remain the onboard optical transmitter and the pointing, acquisition, and tracking (PAT) subsystem, which dominate the size, complexity, and power budgets of existing CubeSat optical terminals \cite{Kingsbury2015,OSIRIS4CubeSat,11071906}.

To reduce onboard complexity, asymmetric link designs have gained interest. These architectures deliberately place most of the complexity of transmission, pointing, and tracking on one side of the link, typically a larger satellite or ground terminal, so that the CubeSat terminal can remain lightweight and power-efficient. For instance, optical downlinks exploit large-aperture ground stations to relax spacecraft transmit requirements \cite{OSIRIS4CubeSat,OCSD}, while LEO–GEO crosslinks have demonstrated multi-gigabit rates using similar asymmetry \cite{LEO_GEO}.  European Space Agency (ESA)’s Hera mission further validated the practicality of asymmetric inter-satellite links with its CubeSat-to-mothership S-band demonstration \cite{Hera}. 

Among the various architectures proposed for CubeSat‑assisted laser communications, the modulating retroreflector (MRR) has emerged as a particularly promising solution due to its simplicity, low power consumption, and suitability for small‑satellite platforms. Retroreflectors, first employed in space for the Lunar Laser Ranging project \cite{LLR}, have since been adapted into MRRs for low-power optical downlinks \cite{Retro_1,Retro_2,Retro_3,Retro_4}. By eliminating the need for an onboard laser transmitter and PAT subsystem, MRRs drastically reduce SWaP overhead, making them highly compatible with CubeSat platforms. Despite these advantages, prior MRR demonstrations have focused almost exclusively on ground-to-space or space-to-ground scenarios, with little attention to inter-satellite communication. More broadly, existing CubeSat optical demonstrations still rely on direct laser transmitters with fine PAT subsystems \cite{Kingsbury2015,OSIRIS4CubeSat,OCSD}, while analytical models often neglect key stochastic impairments such as pointing errors, velocity aberration, and signal-dependent noise. To the best of our knowledge, no prior work has developed a statistical channel model or conducted a comprehensive performance characterization of retroreflector-enabled inter-satellite links under CubeSat SWaP constraints.

Motivated by this gap, we investigate the feasibility of MRR-enabled OISLs for CubeSats. Specifically, we study gigabit-per-second OISLs over distances up to 500 km using on-off keying (OOK) modulation, selected for its simplicity and compatibility with standard optical platforms. Our model explicitly incorporates stochastic pointing losses, deterministic velocity aberration in relay architectures, and signal-dependent noise. Thus, this work presents a statistical channel model and performance analysis for an OOK-modulated, retroreflector-enabled asymmetric inter-satellite link. Based on this framework, we derive analytical expressions for bit-error ratio (BER), outage probability, and achievable information rate (AIR). Using Monte Carlo simulation, we optimize system performance under CubeSat SWaP constraints and benchmark the proposed design against state-of-the-art CubeSat optical terminals. 

The comparison shows that the proposed MRR architecture achieves performance comparable to, and in some cases better than, existing CubeSat optical terminals designed for downlinks, such as the OSIRIS4CubeSat terminal developed by the German Aerospace Center (DLR) \cite{OSIRIS4CubeSat} and NASA’s Optical Communications and Sensors Demonstration (OCSD) program \cite{OCSD}, particularly at short-to-medium ranges below 500 km. Unlike conventional laser terminals, the MRR system is highly tolerant of degree-level CubeSat ADCS pointing errors and operates at a fraction of the power budget, with a maximum draw of only 2.5 W. However, the round-trip geometry inherently imposes a stronger dependence on link length, making velocity aberration and aperture sizing critical for maintaining reliable performance beyond 500 km. These trade-offs indicate that MRR-based OISLs are not intended to replace high-precision crosslink systems such as NASA’s CubeSat Laser Infrared CrosslinK (CLICK) mission \cite{CLICKB_C}, but rather to complement them as a low-SWaP solution well-suited for asymmetric CubeSat communication scenarios.

The remainder of this paper is organized as follows. Section~\ref{sec:retro} reviews retroreflector principles. Section~\ref{sec:model} describes the system model. Section~\ref{sec:results} presents system optimization results. Section~\ref{sec:resource} details the resource budget analysis. Section~\ref{sec:comp} benchmarks our design against existing CubeSat optical terminals. Finally, Section~\ref{sec:conclusion} concludes the paper and discusses future directions. For clarity, we also provide the corresponding abbreviations in Table~\ref{tab:abbreviations}, and a comprehensive list of all symbols and parameters used throughout the paper in Table~\ref{tab:parameters}.


\begin{table}[t]
\centering
\caption{List of Abbreviations}
\begin{tabular}{|l|p{6.2cm}|}
\hline
\textbf{Abbreviation} & \textbf{Full Form / Description} \\
\hline\hline
ADCS & Attitude Determination and Control System \\
AIR & Achievable Information Rate \\
APD & Avalanche Photodiode \\
CCR & Corner-Cube Retroreflector \\
CLICK & CubeSat Laser Infrared CrosslinK \\
EAM & Electro-Absorption Modulator \\
ESA & European Space Agency \\
FEC & Forward Error Correction \\
FFIP & Far-Field Intensity Pattern \\
FOV & Field of View \\
GHz & Gigahertz \\
IM/DD & Intensity Modulation / Direct Detection \\
LEO & Low Earth Orbit \\
LIDAR & Light Detection and Ranging \\
LOS & Line of Sight \\
MEMS & Micro-Electro-Mechanical Systems \\
ML & Maximum Likelihood \\
MQW & Multi-Quantum Well \\
MRR & Modulating Retroreflector \\
NEP & Noise-Equivalent Power \\
OISL & Optical Inter-Satellite Link \\
OOK & On-Off Keying \\
OWC & Optical Wireless Communication \\
PAT & Pointing, Acquisition, and Tracking \\
PDF & Probability Density Function \\
PD & Photodetector \\
PSD & Power Spectral Density \\
RIN & Relative Intensity Noise \\
RF & Radio Frequency \\
Rx & Receiver \\
SNR & Signal-to-Noise Ratio \\
SWaP & Size, Weight, and Power \\
TIA & Transimpedance Amplifier \\
Tx & Transmitter \\
VCSEL & Vertical-Cavity Surface-Emitting Laser \\
\hline
\end{tabular}
\label{tab:abbreviations}
\end{table}

\section{Modulating Retroreflector Principles}\label{sec:retro}

Retroreflectors are optical devices that reflect incident light back along its incoming direction with minimal scattering, even over a wide range of incidence angles. The two most common designs are the corner-cube retroreflector (CCR), which consists of three mutually perpendicular reflective faces, and the cat’s-eye retroreflector, which combines a primary lens with a secondary focal-plane mirror.

An MRR extends this principle by coupling the passive alignment properties of a retroreflector with an active optical modulator \cite{Rabinovich_5}. In this architecture, a conventional laser communication terminal transmits an interrogating beam towards the retroreflector terminal. The beam is retroreflected back after being modulated by the active device, thereby conveying information to the initiator terminal \cite{Rabinovich_1, Booth}.

Different approaches exist for implementing MRR intensity modulation. Liquid crystal (LC) and micro-electro-mechanical systems (MEMS) technologies are suitable for low-bandwidth applications. However, achieving modulation speeds beyond the Megahertz (MHz) range requires surface-normal electro-absorption modulator (EAM) arrays \cite{Quintana,MQW_1,Rabinovich_1}. The performance of EAMs is constrained by their capacitance, as their RC-limited bandwidth scales inversely with device surface area. This poses challenges for CCR-based designs, since efficient modulation requires coverage of the entire entrance aperture. To decouple modulation bandwidth from retroreflector aperture size, cat’s-eye architectures can be combined with focal-plane surface-normal EAM arrays \cite{Rabinovich_4}. Interestingly, such designs have demonstrated Gigahertz (GHz)-class modulation bandwidths \cite{Quintana}. The interrogating beam is a continuous-wave signal, enabling information transfer only in one direction \cite{Rabinovich_5}. Notably, multi-quantum-well (MQW) EAMs exhibit dual functionality as both optical modulators and photodetectors. This property opens avenues for half-duplex optical communication using shared optics and hardware. 


\section{System Model}\label{sec:model}

\begin{table}[t]
\centering
\caption{List of Parameters Used in the Paper}
\begin{tabular}{|l|p{6.2cm}|}
\hline
\textbf{Symbol} & \textbf{Description} \\
\hline\hline
$N$ & Number of Monte Carlo trials \\
$P_t$ & Transmit (interrogator) optical power \\
$\lambda$ & Optical wavelength \\
$B$ & Modulation bandwidth \\
$D_{\text{rr}}$ & Retroreflector diameter \\
$D_{\text{tx}}$ & Transmitter diameter \\
$D_{\text{rx}}$ & Receiver diameter \\
$\theta_{\text{div}}$ & Beam divergence (1/$e^2$ full angle) \\
$\sigma_{\text{tx}}, \sigma_{\text{rx}}$ & Pointing error of transmitter and receiver \\
$\sigma_{\text{rr}}$ & CubeSat post-ADCS pointing error \\
$\text{BER}_{\text{th}}$ & Bit error rate FEC threshold \\
$h_1, h_2$ & CubeSat and interrogator orbital altitudes \\
$\phi$ & Orbital angular separation \\
$\beta_0$ & Optimal aperture-to-beamwaist ratio \\
$\eta_{\text{opt}}$ & System optical efficiency \\
$\eta_{\text{mod}}$ & Modulator insertion loss \\
$N_{\text{EXT}}$ & Modulator extinction ratio \\
$F$ & APD excess noise factor \\
$\theta_{\text{FOV}}$ & Receiver field of view \\
$\Delta\lambda$ & Optical bandpass filter width \\
$F_\lambda$ & Solar spectral irradiance \\
$\Omega_{\text{sol}}$ & Solar solid angle \\
$a$ & Aperture radius \\
$A_{\text{rx}}$ & Receiver aperture area \\
$\eta_{\text{rx}}$ & Receiver optical efficiency \\
$\eta_{\text{tx}}$ & Transmitter optical efficiency \\
$\Omega_{\text{FOV}}$ & Receiver field-of-view solid angle \\
$M$ & Avalanche photodiode (APD) gain \\
$R$ & APD responsivity \\
$I_d$ & APD dark current \\
$P_{\text{sig}}$ & Received optical signal power \\
$P_{\text{bg}}$ & Background optical power \\
$N_{\text{NEP}}$ & Noise-equivalent power term \\
$N_{\text{sh}}$ & Shot noise power spectral density \\
$N_{\text{RIN}}$ & Relative intensity noise PSD \\
$\sigma^2_{\text{N}}$ & Total noise variance \\
$v_\perp$ & Relative transverse velocity \\
$\alpha$ & Round-trip angular offset\\
$L_{\text{rr}}(\alpha,\theta)$ & Retroreflector insertion factor \\
$L_{\text{v}}$ & Velocity aberration loss factor \\
$\Theta_{\text{tx}}, \Theta_{\text{rr}}, \Theta_{\text{rx}}$ & Pointing error random variables \\
$\eta_x$ & Symbol-dependent modulator insertion loss \\
$P_{\text{sig}}(x,g)$ & Signal power for channel state $g$ \\
$i_{\text{rx}}$ & Received photocurrent \\
$I(X;i_{\text{rx}})$ & Achievable information rate (AIR) \\
\hline
\end{tabular}
\label{tab:parameters}
\end{table}

\begin{figure*}[t]
    \centering
    \includegraphics[height=7.0cm]{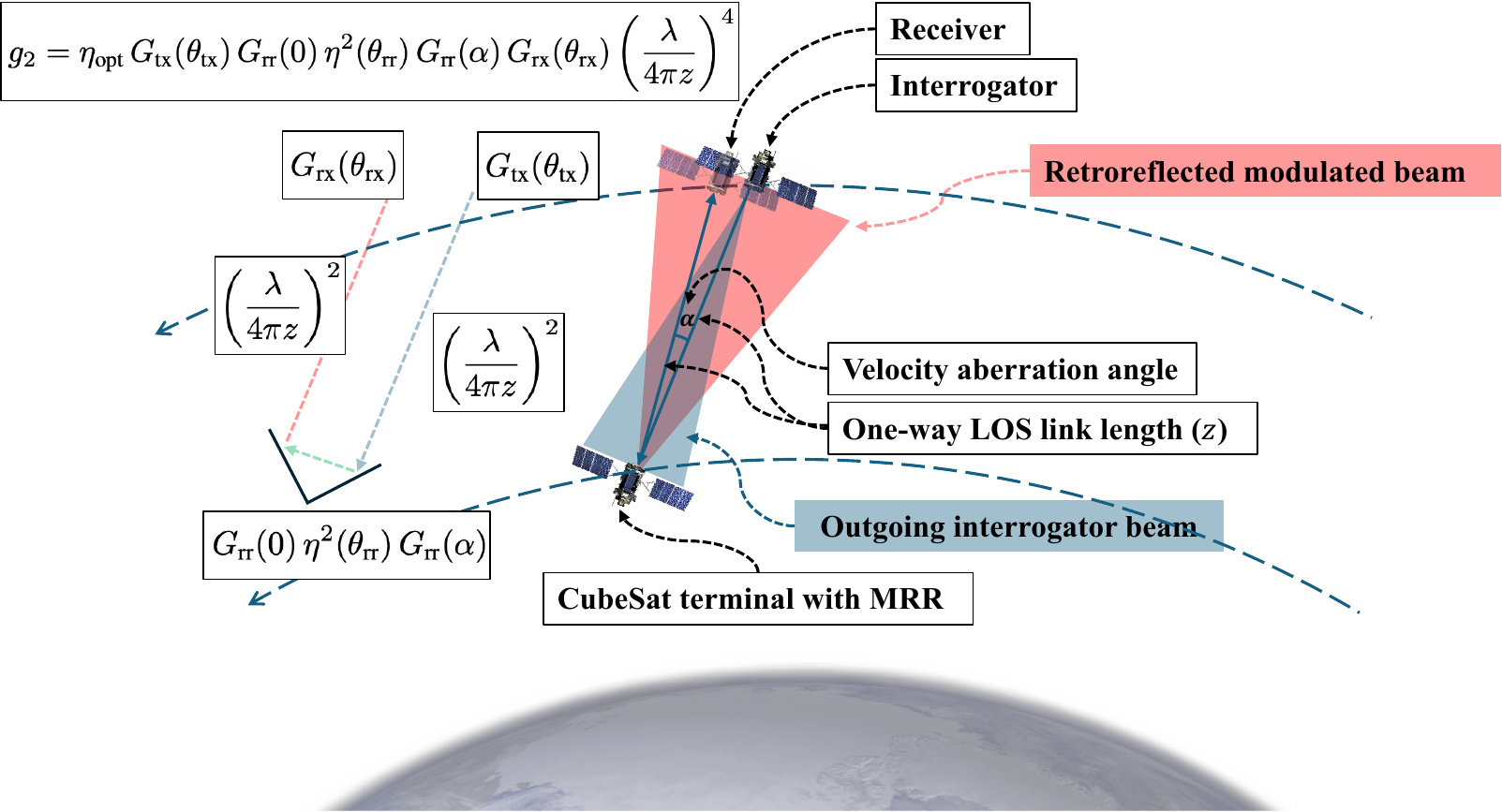} 
    \caption{System model illustrating the round-trip communication link setup between two satellites and key parameters influencing the channel characteristics and overall system performance.}
    \label{fig:sysmodel}
\end{figure*}

A schematic of the round-trip retroreflector-based optical inter-satellite link is shown in Fig.~\ref{fig:sysmodel}, and the corresponding parameters are listed in Table~\ref{tab:parameters}. \textcolor{black}{Unless otherwise stated, numerical examples throughout this section use the parameters listed in Table~\ref{tab:Example Parameters}.} The system comprises an active interrogator terminal that transmits an optical beam to a secondary terminal equipped with an MRR. The following subsections first formulate a one-way link model and subsequently extend the model to the full round-trip channel.

\begin{table}[h]
\centering
\caption{\textcolor{black}{Link Parameters Used in Section~\ref{sec:model}}}
\label{tab:Example Parameters}
\begin{tabular}{|l|l|l|}
\hline
\textbf{Name} & \textbf{Parameter} & \textbf{Value} \\
\hline\hline
Number of trials & $N$ & $1\times10^7$ \\
Orbital angular separation & $\phi$ & $0^{\circ}$\\
Optical Wavelength & $\lambda$ & 850 nm \\
One-way LOS Link length & $z$ & 1000 km \\
Transmitter Diameter & $D_{\text{tx}}$ & 10 cm \\
Receiver Diameter & $D_{\text{rx}}$ & 10 cm \\
Retroreflector Diameter & $D_{\text{rr}}$ & 10 cm \\
Receiver field of view & $\theta_{\text{FOV}}$ & 100 $\mu\text{rad}$\\
Pointing error (Tx/Rx) & $\sigma_{tx}, \sigma_{rx}$ & 1 $\mu$rad \\
CubeSat pointing error & $\sigma_{\text{rr}}$ & $1^\circ$ \\
\hline
\end{tabular}
\end{table}

\subsection{Point-to-point Optical Transmission}\label{sec:p2p}

The starting point for modeling free-space optical propagation is the Friis transmission equation \cite{ITU_R_S.1590_2002}. We extend this formulation to optical frequencies using Fourier optics, representing a diffraction-limited aperture as an effective gain factor. This gain is defined as the ratio between the far-field intensity produced by the illuminated aperture and that of an isotropic radiator of equal power \cite{Klein1974-uc}.

The far-field intensity pattern (FFIP) depends on the aperture illumination profile. Two cases are of particular interest, i.e., uniform (plane) and Gaussian illumination. For an aperture of area $A$ and wavelength $\lambda$, the directional gain is expressed as
\begin{equation}
\label{eq:01}
G_a(\theta) = \left(\frac{4\pi A}{\lambda^2}\right) L(\theta),
\end{equation}
where
\begin{enumerate}
    \item [-] $\theta$ is the boresight angle defined as the angle between the optical axis of the transmitting aperture and the line of sight from the transmitter aperture center to the detector location in the far field.
    \item [-] $L(\theta)$ characterizes the angular profile of the far-field intensity pattern, as determined by the aperture illumination profile.
\end{enumerate}

For a uniformly illuminated circular aperture of radius $a$ and wavenumber $k$, the FFIP, denoted $L_{\text{u}}$, follows the Airy pattern, thus we have \cite{Degnan2023-bz}
\begin{equation}
\label{eq:02}
L_{\text{u}}(\theta) = \left|2\frac{J_1(X)}{X}\right|^2_{X = ka\sin{\theta}},
\end{equation}
where $J_1(\cdot)$ is the Bessel function of the first kind of order one.

\begin{figure}[h]
\centering
\includegraphics[width=0.9\columnwidth]{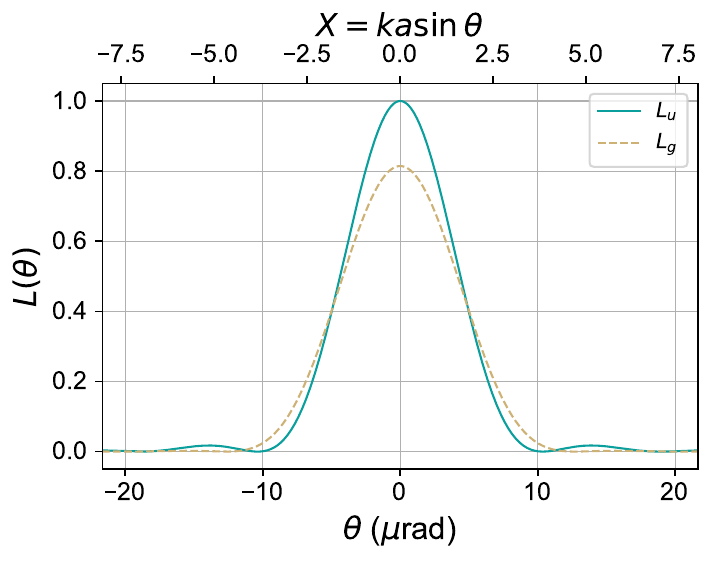}
\caption{\textcolor{black}{FFIP for uniform ($L_{\text{u}}$) and Gaussian ($L_{\text{g}}$) beams plotted as 
functions of the off-boresight angle $\theta$ using the dimensionless angular 
coordinate $X = ka\sin\theta$. For the parameters in 
Table~\ref{tab:Example Parameters}, uniform illumination yields a divergence of 
$\sim$10~$\mu$rad with unit peak gain, while Gaussian illumination produces a 
broader divergence of $\sim$12.5~$\mu$rad and a reduced on-axis peak of $\sim$0.8.}} 
\label{fig:L_g}
\end{figure}

Also, for a Gaussian beam with beam waist $\omega$ truncated by the same circular aperture, the FFIP, denoted $L_g$, is expressed as \cite{Klein1974-uc}
\begin{equation}
\label{eq:03}
L_{\text{g}}(\theta) = 2\beta_0^2 \left|\int_0^1\exp{\left(-\beta_0^2u\right)}J_0\left(X\sqrt{u}\right)du\right|^2_{X = ka\sin{\theta}},
\end{equation}
where
\begin{enumerate}
    \item [-] $J_0(\cdot)$ denotes the Bessel function of the first kind of order zero.
    \item [-] $\beta_0$ represents the optimal ratio between the aperture radius and the beam waist that maximizes the on-axis gain $L(0)$. $\beta_0 = 1.12$ \cite{Klein1974-uc}.
\end{enumerate}
\textcolor{black}{Fig. \ref{fig:L_g} illustrates the normalized FFIPs $L_{\text{u}}(\theta)$ and $L_{\text{g}}(\theta)$ plotted as functions of the off-boresight angle $\theta$, expressed in terms of the normalized angular coordinate $X = ka\sin\theta$, as defined in (\ref{eq:02})–(\ref{eq:03}).} \textcolor{black}{For a given illumination profile and a circular aperture of radius $a$ (diameter $D = 2a$), the diffraction-limited divergence angle, commonly defined by the first null of the angular response, scales as $\theta_{\text{div}} \propto \lambda/D = 1/(ka)$. Under the small-angle approximation $\sin\theta \approx \theta$, the variable $X$ therefore represents the off-axis angle normalized by the aperture’s diffraction divergence, i.e., $X \propto \theta/\theta_{\text{div}}$. A larger divergence angle, corresponding to a smaller transmit aperture or longer wavelength, permits larger angular deviations before the far-field response decays.}   

The transmit gain $G_{\text{tx}}(\theta)$ represents the directional gain of the laser transmitter operating in the fundamental Gaussian mode (TEM$_{00}$). It follows from (\ref{eq:01}) as
\begin{equation}
\begin{split}
    G_{\text{tx}}(\theta) &= \left(\frac{4\pi A_{\text{tx}}}{\lambda^2}\right)L_{\text{tx}}(\theta)\\
    &=\left(\frac{4\pi A_{\text{tx}}}{\lambda^2}\right)L_{\text{g}}(\theta)
\end{split}
\end{equation}
where $A_{\text{tx}}$ is the transmitter aperture area, and $L_{\text{tx}}(\theta) = L_{\text{g}}(\theta)$ describes the far-field intensity distribution of a truncated Gaussian beam according to (\ref{eq:03}).

\begin{figure}[h]
\centering
\includegraphics[width=0.9\columnwidth]{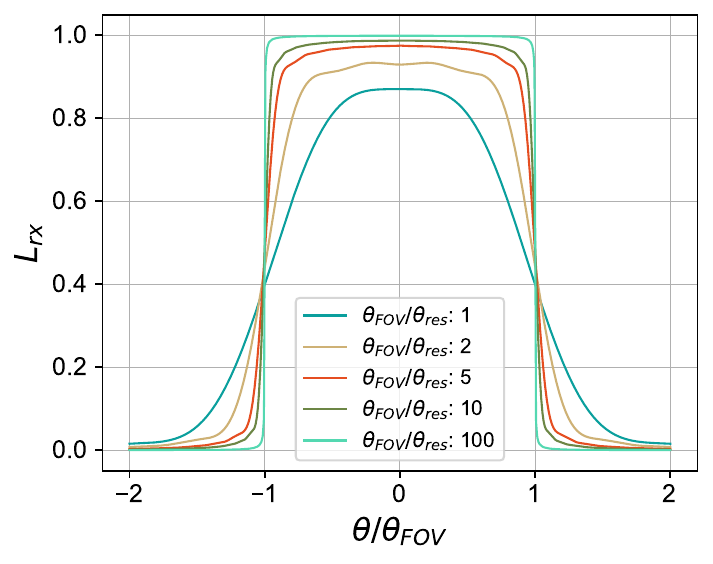}
\caption{\textcolor{black}{Receiver gain profiles for varying $\theta_{\text{FOV}}/\theta_{\text{res}}$, normalized to the uniform-aperture on-axis gain.}}
\label{fig:L_pd}
\end{figure}

\textcolor{black}{Assuming that the received beam footprint is much larger than the receiver aperture, the illumination across the aperture can be approximated as locally uniform, even though the overall beam exhibits the Gaussian far field intensity profile given in \eqref{eq:03} \cite{Safi:20, dabiri2022}.} At the receiver, however, the effective gain also depends on the finite photodetector area $S_{\text{pd}}$. We define $l_{\text{u}}(x, y)$ as the area-normalized focal-plane intensity distribution 
corresponding to the far-field pattern $L_{\text{u}}(\theta)$:
\begin{equation}
\label{eq:lu_def}
l_{\text{u}}(x, y) \propto L_{\text{u}}\!\bigl(\theta(x,y)\bigr),
\end{equation}
normalized such that
\[
\iint l_{\text{u}}(x, y)\,dx\,dy = 1,
\]
where $f$ is the focal length, and $\theta(x,y) \simeq \sqrt{(x/f)^2 + (y/f)^2}$ under the small-angle approximation.
Also, assuming uniform responsivity across the detector, the off-axis gain, $L_{\text{rx}}$, is defined as
\begin{equation}
\label{eq:05a}
L_{\text{rx}}(\theta) = \iint_{S_{\text{pd}}} l_{\text{u}}(x- x_{\theta},y-y_{\theta})dxdy,
\end{equation}
where $(x_{\theta},y_{\theta})$ denotes the focal-plane coordinates of the beam center corresponding to an off-axis angle $\theta$.
The resulting gain profile is then shaped by the ratio of the photodetector’s angular field of view $\theta_{\text{FOV}}$ to the aperture’s angular resolution $\theta_{\text{res}}$. 
Here, $\theta_{\text{FOV}} \simeq r_{\mathrm{pd}}/f$ denotes the half-angle subtended by a circular photodetector of radius $r_{\mathrm{pd}}$ as seen from the aperture focal point. Parameter $\theta_{\text{res}} \simeq 1.22\,\lambda / D_{\text{rx}}$ represents the diffraction-limited angular resolution of the receiving aperture.
Following the formulation in (\ref{eq:01}), the receiver gain can be expressed as
\begin{equation}
G_{\text{rx}}(\theta) = \left(\frac{4\pi A_{\text{rx}}}{\lambda^2}\right)L_{\text{rx}}(\theta),
\end{equation}
where $A_{\text{rx}}$ is the receiver aperture area, and $L_{\text{rx}}(\theta)$ describes the angular gain profile of the receiving aperture with a finite photodetector, as defined in (\ref{eq:05a}). The corresponding gain profiles are depicted in Fig. \ref{fig:L_pd}. 

\textcolor{black}{For $\theta_{\text{FOV}}/\theta_{\text{res}} =1$, the profile has a flattened central peak, with a gradual roll-off. $L_{\text{rx}}(0) < 1$ reflecting partial truncation of the diffraction-limited Airy pattern by the photodetector area. As $\theta_{\text{FOV}}/\theta_{\text{res}}$ increases, a larger fraction of the Airy disk is captured by the photodetector, and $L_{\text{rx}}$ approaches a top-hat response with a transition near $\theta_{\text{FOV}}$. In this limit, the on-axis value $L_{\text{rx}}(0)$ approaches unity, indicating near-complete capture of the diffraction-limited spot on the focal plane.}

By accounting for both transmit- and receive-side effects, 
the deterministic channel gain for a given pair of angular misalignments $(\theta_{\text{tx}}, \theta_{\text{rx}})$ 
can be expressed as
\begin{equation}
\label{eq:g}
\begin{split}
      g_1 &= \eta_{\text{tx}}\eta_{\text{rx}}\,G_{\text{tx}}\!\left(\theta_{\text{tx}}\right)
      G_{\text{rx}}\!\left(\theta_{\text{rx}}\right)
      \left(\frac{\lambda}{4\pi z}\right)^{2}\\
      &=
      \begin{split}
          &\left(\frac{4\pi A_{\text{tx}}}{\lambda^{2}}\right)
          \left(\frac{4\pi A_{\text{rx}}}{\lambda^{2}}\right)
          \left(\frac{\lambda}{4\pi z}\right)^{2}\\
          &\times
          \left[\eta_{\text{tx}}\eta_{\text{rx}}L_{\text{tx}}\!\left(\theta_{\text{tx}}\right)
          L_{\text{rx}}\!\left(\theta_{\text{rx}}\right)\right],
      \end{split}
\end{split}
\end{equation}
where $\eta_{\text{tx}}$ and $\eta_{\text{rx}}$ accounts for efficiencies at the transmitter and receiver optical chains, respectively. $A_{\text{tx}}$ and $A_{\text{rx}}$ are the aperture areas, 
and $L_{\text{tx}}(\cdot)$ and $L_{\text{rx}}(\cdot)$ represent the off-axis gain factors at the transmitter and receiver, respectively.

\begin{figure}[h]
\centering
\includegraphics[width=0.9\columnwidth]{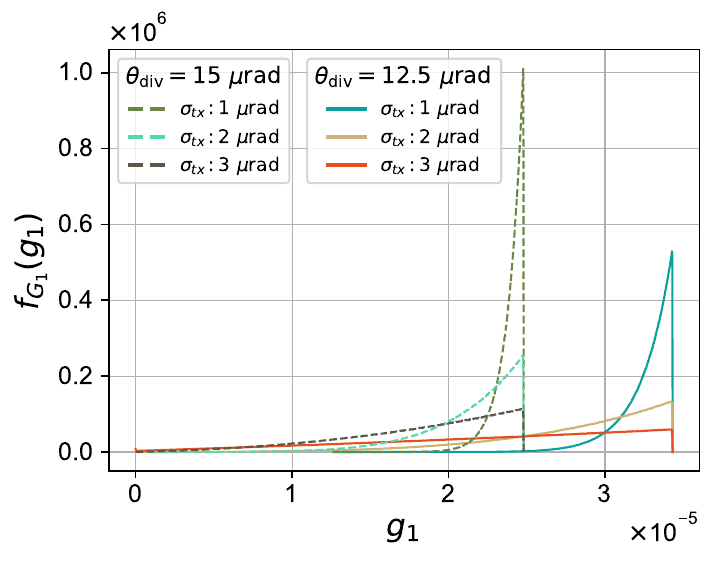}
\caption{\textcolor{black}{Distribution of channel gain $f_{G_1}(g_1)$ vs. ratio of transmit divergence angle to pointing error. Larger ratios concentrate probability near the on-axis maximum.}}
\label{fig:g_tx}
\end{figure}

\begin{figure}[h]
\centering
\includegraphics[width=0.9\columnwidth]{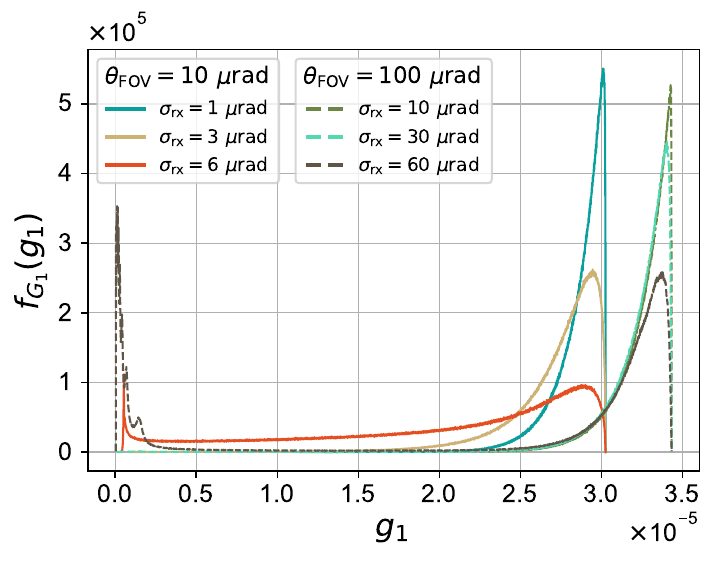}
\caption{\textcolor{black}{Distribution of channel gain $f_{G_1}(g_1)$ vs. receiver FOV. Wider FOVs increase gain.}}
\label{fig:g_rx}
\end{figure}

In practical laser communication systems, pointing errors have both deterministic and stochastic components. The deterministic component arises from position determination errors and inaccuracies in boresight calibration, while the stochastic component is caused by residual uncompensated platform jitter \cite{HEMMATI2013121}.
In well-calibrated systems with accurate point-ahead error compensation, 
residual pointing errors can be modeled as independent zero-mean Gaussian fluctuations along two orthogonal axes about the nominal line-of-sight, 
resulting in Rayleigh-distributed off-axis angles \cite{Safi:20}. Consequently, the transmitter and receiver misalignment angles, $\Theta_{\text{tx}}$ and $\Theta_{\text{rx}}$, are modeled as Rayleigh-distributed random variables with single-axis standard deviations $\sigma_{\text{tx}}$ and $\sigma_{\text{rx}}$ as follows
\begin{equation}
\label{eq:theta_tx}
    \Theta_{\text{tx}} \sim \text{Rayleigh}(\sigma_{\text{tx}}),
\end{equation}
\begin{equation}
\label{eq:theta_rx}
    \Theta_{\text{rx}} \sim \text{Rayleigh}(\sigma_{\text{rx}}).
\end{equation}
As a result, the end-to-end channel gain becomes a random variable, with a distribution shown in Figs. \ref{fig:g_tx}--\ref{fig:g_rx}, and expressed as
\begin{equation}
G_1 \sim \mathcal{G}_1(\sigma_{\text{tx}}, \sigma_{\text{rx}}).
\end{equation}


Although a closed form expression for the distribution is not available, numerical evaluation provides insight into how pointing errors influence the one way channel gain $G_1$. As shown in Figs. \ref{fig:g_tx} and \ref{fig:g_rx}, the distribution is primarily governed by the ratios $\sigma_{\text{tx}}/\theta_{\text{div}}$, $\sigma_{\text{rx}}/\theta_{\text{FOV}}$ and $\theta_{\text{FOV}}/\theta_{\text{res}}$. Reducing $\sigma_{\text{tx}}$ and $\sigma_{\text{rx}}$ relative to $\theta_{\text{div}}$ and $\theta_{\text{FOV}}$, respectively, concentrates the probability mass near the on axis condition, since 99.7\% of Rayleigh distributed pointing errors lie within three standard deviations. 

Increasing the transmit divergence $\theta_{\text{div}}$ improves robustness to transmit-side pointing errors but reduces on-axis gain, which scales as $G_{\mathrm{tx}}(0) \propto 1/\theta_{\mathrm{div}}^2$. On the receive side, increasing $\theta_{\mathrm{fov}}$ reduces susceptibility to pointing errors, and can increase on-axis gain through $L_{\mathrm{rx}}$, if increased relative to $\theta_{\text{res}}$. Reducing $\theta_{\text{res}}$ increases on-axis gain through $L_{\mathrm{rx}}$ when $\theta_{\text{FOV}}/\theta_{\text{res}}$ is increased. For fixed $\theta_{\text{FOV}}/\theta_{\text{res}}$, reducing $\theta_{\text{res}}$ increases on-axis gain as $G_{\mathrm{rx}}(0) \propto 1/\theta_{\mathrm{res}}^2$. 

An optimal system trades a narrow beam divergence for resilience against pointing errors, maximizes receiver aperture for on-axis gain, and balances a wide receiver FOV for jitter tolerance against increased susceptibility to background noise, as discussed further in Section~\ref{sec:results}.

\subsection{Retroreflector-based Round-trip Model}

The Friis propagation model can be naturally extended to retroreflector-based links by modeling the retroreflector as operating jointly in receive and transmit modes. Under this formulation, the round-trip channel gain is represented as the cascade of two point-to-point propagation paths \cite{dabiri2022}. A key factor in this round-trip gain is the retroreflection efficiency, defined as the fraction of incident power returned along the incoming direction. 
\textcolor{black}{For a hollow CCR illuminated at incidence angle $\theta$ relative to the aperture normal, the lateral separation between the incident and reflected beams at the aperture plane is obtained as}
\begin{equation}
\label{eq:d}
    d = 2L\tan\theta,
\end{equation}
where $L$ is the distance from the cube vertex to the aperture plane~\cite{1979SAOSR.382.....A}.
Subsequently, this geometric relationship, together with the aperture shape, determines the portion of the aperture that remains active for retroreflection. For a circular aperture of radius $a_{\text{rr}}$, the effective retroreflecting area is~\cite{1979SAOSR.382.....A}

\begin{equation}
\label{eq:06}
    A_{\text{eff}} = 2a_{\text{rr}}^2(\psi - \cos{\psi}\sin{\psi}),
\end{equation}
where $\cos{\psi} =\frac{L}{a_{\text{rr}}}\tan{\theta}$, and $\frac{L}{a_{\text{rr}}}\tan{\theta} < 1$ for retroreflection to occur.
Accordingly, the retroreflection efficiency $\eta$ is defined as the ratio between the effective aperture area projected along the incidence direction and the physical aperture area, and can be expressed as
\begin{equation}
\label{eq:07}
    \eta(\theta) = \left\{\begin{array}{ll}
    \frac{2}{\pi}(\psi - \cos{\psi}\sin{\psi})\cos{\theta}& 0<\tan{\theta }< {\frac{a}{L}},\\
    0 & \text{otherwise}.\\
\end{array}
\right.
\end{equation}
Under the paraxial and thin-lens approximations, the lateral separation between the incident and reflected rays in a cat’s-eye retroreflector is given by
\begin{equation}
\label{eq:cats_eye_d}
    d = 2f\tan\theta,
\end{equation}
where $f$ is the lens focal length and $\theta$ is the incidence angle relative to the optical axis.
Moreover, by analogy to the corner-cube case, the incidence-dependent efficiency for a cat’s-eye retroreflector with focal length $f$ is given by
\begin{equation}
\label{eq:08}
    \eta(\theta) = \left\{\begin{array}{ll}
    \frac{2}{\pi}(\psi^{'} - \cos{\psi^{'}}\sin{\psi^{'}})\cos{\theta}& 0<\tan{\theta }< {\frac{a_{\text{rr}}}{f}},\\
    0 & \text{otherwise}.\\
\end{array}
\right.
\end{equation}
where $\cos{\psi^{'}} =\frac{f}{a_{\text{rr}}}\tan{\theta}$.

Assuming that the received beam footprint is much larger than the retroreflector aperture, the aperture illumination can be approximated as locally uniform. For small incidence angles, the effective retroreflector area remains approximately circular and diffraction limited conditions apply. Under these assumptions, the far field intensity pattern is well approximated by an Airy profile \cite{Degnan2023-bz}. Accordingly, the retroreflector gain follows the forms in \eqref{eq:01} and \eqref{eq:02}, with the aperture area in \eqref{eq:01} scaled by the incidence dependent efficiency $\eta$, and the aperture radius in \eqref{eq:02} scaled by $\eta^{1/2}$.

By construction, the retroreflector’s receive axis is normal to the effective projected aperture, and thus aligned with the incident beam. Hence, the retroreflector experiences on-axis gain in the receive mode. The transmit gain, however, is affected by relative transverse motion between the terminals. Due to the finite speed of light, this motion introduces a systematic angular offset between the outgoing and returning beams. As a result,  the retroreflected beam is received at angle $\alpha$ relative to the transmit axis (as shown in Fig. \ref{fig:sysmodel}). This velocity aberration reduces the effective gain \cite{Degnan2023-bz}. Considering the round-trip offset, $\alpha$ (measured in radians) is given by 
\begin{equation}
\label{eq:10}
\alpha = \frac{2v_{\perp}}{c},
\end{equation}
where $v_{\perp}$ is the relative transverse velocity between the transmitter and retroreflector, and $c$ is the speed of light. Note that the transverse velocity $v_{\perp}$, and consequently the angular offset $\alpha$, are determined by orbital mechanics. In this work, we consider a generic optical relay scenario involving two satellites in circular orbits at different altitudes and with angularly separated orbital planes. A full derivation of the resulting relative motion is provided in the Appendix.

Indeed, velocity aberration manifests as a systematic pointing loss, i.e., the retroreflected beam is shifted off-axis by angle $\alpha$, reducing the effective return gain. This aberration-induced loss is
\begin{equation}
\label{e:10}
L_\text{v}(\alpha) = \left|2\frac{J_1(X)}{X}\right|^2_{X = ka_{\text{rr}}\sin{\alpha}}.
\end{equation}
Including both incidence efficiency and velocity aberration yields the total retroreflector insertion factor as follows
\begin{equation}
\label{e:11}
L_{\text{rr}}(\alpha,\theta) = \eta(\theta)^2\left|2\frac{J_1(X)}{X}\right|^2_{X = \eta(\theta)^{1/2}ka_{\text{rr}}\sin{\alpha}}.
\end{equation}
\begin{figure*}[t]
\centering
\includegraphics[width=\textwidth]{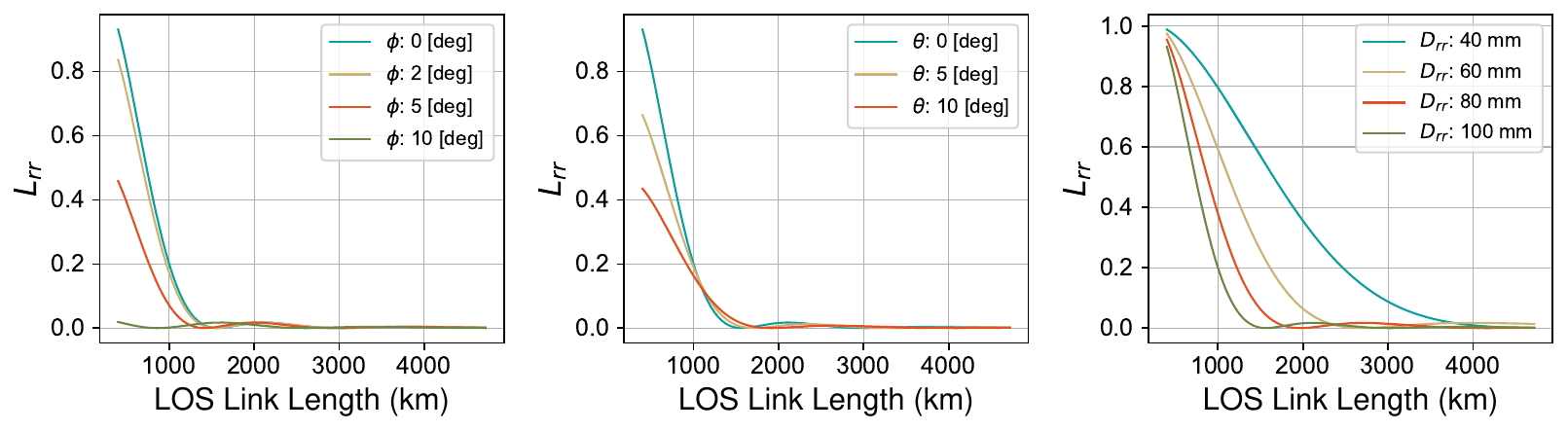}
\caption{Retroreflector insertion factor $L_{\text{rr}}$ versus LOS link length at 400 km separation: (A) varying orbital angular separation, $\phi$, at $\theta=0$, and $D_{\text{rr}} = 100\text{ mm}$ (B) varying incidence angle, $\theta$, at $\phi = 0$, and $D_{\text{rr}} = 100\text{ mm}$  (C) varying aperture diameter and $D_{\text{rr}}$  at $\theta=0$ and $\phi=0$.}
\label{fig:L_rr}
\end{figure*}
%
The dependence of $L_{\text{rr}}$ on link geometry is shown in Fig.~\ref{fig:L_rr}. As the angular separation $\phi$ between the satellites’ orbital planes increases, the higher transverse velocity leads to larger $\alpha$, driving the return beam toward the Airy sidelobes. For the chosen parameters, significant loss appears at orbital separations above $\phi>2^{\circ}$, and the maximum feasible link length is limited to $\sim1000$ km, before $\alpha$ reaches the first Airy null (see Fig.~\ref{fig:L_rr}(a)).
Also, incidence angle $\theta$  affects performance through the efficiency term $\eta(\theta)$. As incidence increases, the effective aperture shrinks by $\eta^{1/2}$, widening the divergence angle by the same factor but reducing the on-axis gain by $\eta^2$. Fig.~\ref{fig:L_rr}(b) shows that this trade-off benefits only very long links ($>1000$ km), where $L_{\text{rr}}$ is already degraded. For practical OISLs, maintaining strong on-axis gain is preferable to trading efficiency for divergence.
Furthermore, aperture size introduces another trade-off. Since the Airy divergence angle scales inversely with aperture diameter $D_{\text{rr}} = 2a_{\text{rr}}$, reducing $D_{\text{rr}}$ broadens the beam and allows tolerance to larger $\alpha$ before significant misalignment occurs. However, the round-trip link gain scales as $D_{\text{rr}}^4$. Even modest reductions in $D_{\text{rr}}$ therefore impose severe on-axis penalties, offsetting any gain from improved aberration tolerance as shown in Fig.~\ref{fig:L_rr}(c).

\begin{figure}[h]
\centering
\includegraphics[width=0.9\columnwidth]{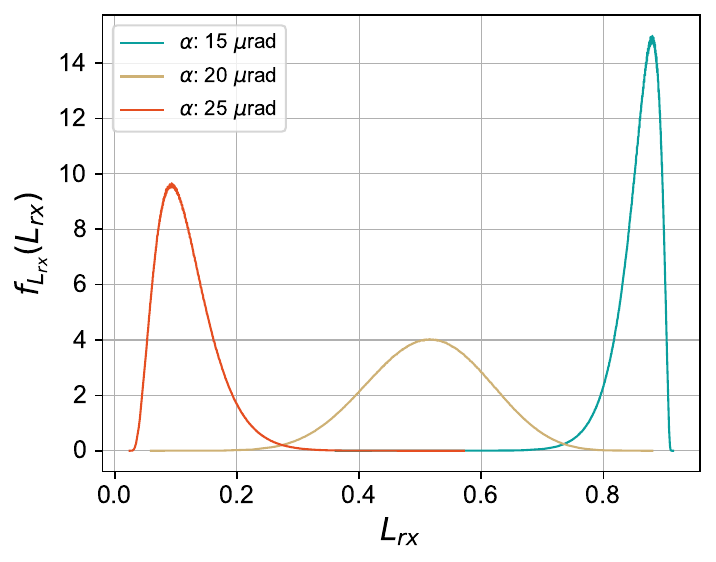}
\caption{\textcolor{black}{Normalized distribution of receiver gain $L_{\text{rx}}$ under systematic angular offsets $\alpha$. Increasing $\alpha$ shifts probability mass away from on-axis gain, reducing performance. Example shown for $\theta_{\text{FOV}}=20~\mu$rad and parameters in Table~\ref{tab:Example Parameters}.}}
\label{fig:L_pd_2}
\end{figure}

We define the random pointing errors associated with the three terminals as $\Theta_{\text{tx}}$, $\Theta_{\text{rr}}$, and $\Theta_{\text{rx}}$, corresponding to the interrogator, retroreflector, and receiver, respectively. These pointing errors are modeled in an analogous manner to the one-way case described in Section~\ref{sec:p2p}. Assuming zero systematic one-way point-ahead error, the interrogator and retroreflector misalignment angles follow Rayleigh distributions with single-axis standard deviations $\sigma_{\text{tx}}$ and $\sigma_{\text{rr}}$
\begin{equation}
\label{eq:11}
    \Theta_{\text{tx}} \sim \text{Rayleigh}(\sigma_{\text{tx}}),
\end{equation}
\begin{equation}
\label{eq:12}
    \Theta_{\text{rr}} \sim \text{Rayleigh}(\sigma_{\text{rr}}).
\end{equation}
Relative motion impacts not only the mean on-axis gain through velocity aberration but also the statistics of the receiver pointing error. Assuming \textcolor{black}{co-boresighted transmit and receive optics}, the systematic angular offset $\alpha$ \textcolor{black}{biases the receiver pointing angle, shifting its distribution from Rayleigh to Rician. Thus, $\Theta_{\text{rx}}$ follows a nonzero-mean bivariate Gaussian with single-axis standard deviation $\sigma_{\text{rx}}$
\begin{equation}
\label{eq:13}
    \Theta_{\text{rx}} \sim \text{Rice}(\alpha,\sigma_{\text{rx}}).
\end{equation}}
We evaluate the impact of this bias on the receiver gain profile $L_{\text{rx}}$ in Fig.~\ref{fig:L_pd_2}. For small offsets (i.e., when $\alpha < \theta_{\text{FOV}}$), the induced distribution closely resembles a reflected Rayleigh profile, peaking near the maximum value with a shallow left tail. As $\alpha$ increases, the distribution shifts away from the maximum value, corresponding to a systematic reduction in link performance. In addition to $\alpha$, the receiver gain also depends on random jitter $\sigma_{\text{rx}}$ and the receiver field of view $\theta_{\text{FOV}}$. As shown in the one-way case of Fig.~\ref{fig:g_rx}, larger $\sigma_{\text{rx}}$, relative to $\theta_{\text{FOV}}$ broadens the pointing distribution and flattens the gain profile, further degrading link robustness.

Finally,  the deterministic round-trip channel gain for a given set of angular misalignments 
$(\theta_{\text{tx}}, \theta_{\text{rr}}, \theta_{\text{rx}})$ and systematic offset $\alpha$ 
can be expressed as
\begin{equation}
\label{eq:g2}
\begin{split}
      g_2 &= \eta_{\text{opt}}\,G_{\text{tx}}\!\left(\theta_{\text{tx}}\right)
      G_{\text{rr}}\!\left(0\right)
      \eta^{2}\!\left(\theta_{\text{rr}}\right)
      G_{\text{rr}}\!\left(\alpha\right)
      G_{\text{rx}}\!\left(\theta_{\text{rx}}\right)
      \left(\frac{\lambda}{4\pi z}\right)^{4}\\
      &=
      \begin{split}
          &\left(\frac{4\pi A_{\text{tx}}}{\lambda^{2}}\right)
          \left(\frac{4\pi A_{\text{rr}}}{\lambda^{2}}\right)^{2}
          \left(\frac{4\pi A_{\text{rx}}}{\lambda^{2}}\right)
          \left(\frac{\lambda}{4\pi z}\right)^{4}\\
          &\times
          \left[\eta_{\text{opt}}\,
          L_{\text{tx}}\!\left(\theta_{\text{tx}}\right)
          L_{\text{rr}}\!\left(\alpha,\theta_{\text{rr}}\right)
          L_{\text{rx}}\!\left(\theta_{\text{rx}}\right)\right],
      \end{split}
\end{split}
\end{equation}
where $\eta_{\text{opt}}= \eta_{\text{tx}}\eta_{\text{rr}}\eta_{\text{rx}}$. Furthermore, parameters  
$A_{\text{tx}}$, $A_{\text{rr}}$, and $A_{\text{rx}}$ are the physical aperture areas, and $L_{\text{tx}}(\cdot)$, $L_{\text{rr}}(\cdot)$, and $L_{\text{rx}}(\cdot)$ represent the off-axis gain factors at the transmitter, retroreflector, and receiver, respectively. 

\begin{figure}[h]
\begin{center}
\includegraphics[width=0.9\columnwidth]{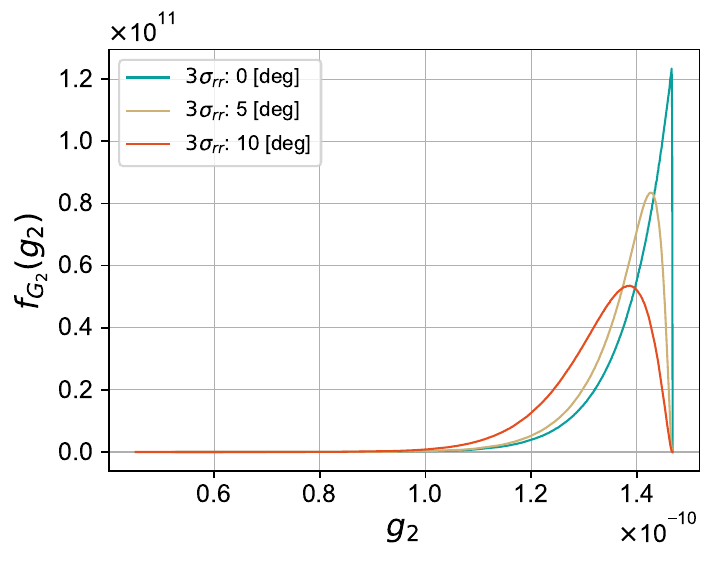}

\end{center}
\caption{\textcolor{black}{Probability distribution of the round-trip MRR channel gain $f_{G_2}(g_2)$ in \eqref{eq:G2} for different CubeSat post-ADCS 3--$\sigma$ pointing errors ($\sigma_{\text{rr}}$).}}
\label{fig:g distribution-sigma_rr}
\end{figure}

In the presence of stochastic pointing errors, whose distributions are given in \eqref{eq:11}–\eqref{eq:13}, the end-to-end round-trip channel gain becomes a random variable,
\begin{equation}
\label{eq:G2}
G_2 = g_2(\Theta_{\text{tx}}, \Theta_{\text{rr}}, \Theta_{\text{rx}}; \alpha),
\end{equation}
whose statistics can be compactly represented as
\begin{equation}
G_2 \sim \mathcal{G}2(\sigma_{\text{tx}}, \sigma_{\text{rr}}, \sigma_{\text{rx}}; \alpha).
\end{equation}
The resulting probability density function $f_{G_2}(g_2)$ does not admit a closed form expression and is therefore evaluated numerically using Monte Carlo simulation. Fig.~\ref{fig:g distribution-sigma_rr} illustrates the impact of retroreflector jitter on the channel gain distribution. As the post ADCS pointing error $\sigma_{\text{rr}}$ increases, the distribution becomes broader and its peak shifts toward lower gain values, indicating both a reduction in the mean channel gain and an increased likelihood of deep fades.

\subsection{Modulation and Receiver Noise Model}

In both direct-transmission and modulating-retroreflector links, modulation is achieved using an electro-absorption modulator (EAM), characterized by its on-state insertion loss $\eta_{\text{mod}}$ and extinction ratio $N_{\text{EXT}}$. The extinction ratio quantifies the modulator’s ability to distinguish between the logical “$1$” and “$0$” states, defined as the ratio of transmitted optical power in the on-state to that in the off-state. A higher extinction ratio improves symbol distinguishability but typically comes at the cost of increased insertion loss and power consumption. In the off-state, the effective insertion loss rises to $\eta_{\text{mod}}/N_{\text{EXT}}$. Thus, one can write the symbol-dependent insertion loss $\eta_x$ for OOK symbols $x \in \{0,1\}$ as

\begin{equation}
\eta_0 = \frac{\eta_{\text{mod}}}{N_{\text{EXT}}}, \qquad \eta_1 = \eta_{\text{mod}}.
\end{equation}
The received signal power for a given instantaneous channel gain \textcolor{black}{$g\in \{g_1,g_2\}$} is then expressed as
\textcolor{black}{\begin{equation}
\label{e:14}
P_{\text{sig}}(x,g) = \eta_x P_{\text{tx}} g,
\end{equation}
where $P_{\text{tx}}$ denotes the laser transmit power.}

The receiver employs direct detection of the transmitted beam intensity. A photodetector (PD) converts incident optical power to a baseband photocurrent, which is amplified by a transimpedance amplifier (TIA) before digitization. For an avalanche PD (APD) with gain $M$, responsivity $R$, and dark current $I_{\text{d}}$, the mean photocurrent generated in response to both the desired signal $P_{\text{sig}}$ and background illumination $P_{\text{bg}}$ is obtained as \cite{Ghassemlooy2019-kk}
\begin{equation}
\label{eq:15}
\langle i_{\text{rx}}(x,g)\rangle = M R\left(P_{\text{sig}}(x,g) + P_{\text{bg}}\right) + I_{\text{d}}.
\end{equation}
Here, the expectation $\langle \cdot\rangle$ is taken over noise realizations only, for fixed transmitted symbol $x$ and channel realization $g$.

 Illumination-independent contributions, including thermal noise, dark current shot noise, and amplifier noise, define the noise floor. This contribution is represented by the input referred transimpedance amplifier noise power spectral density, denoted $N_{\text{floor}}$, which can be expressed in terms of the noise equivalent power (NEP) as \cite{ThorlabsNEP}
\begin{equation}
\label{eq:16}
N_{\text{floor}} = M^{2} R^{2}  \mathrm{NEP}.
\end{equation}
In addition, illumination-dependent noise components include signal-dependent shot noise, background shot noise, and laser relative intensity noise (RIN).
In LEO optical links, the primary sources of background illumination are
\begin{itemize}
    \item Direct solar radiation,
    \item Solar radiation reflected from the Earth’s surface (albedo), and
    \item Internal optical reflections (crosstalk) between the laser transmitter (interrogator) and the photodetector.
\end{itemize}
In this study, the effect of crosstalk is neglected as it can be substantially mitigated through optical isolation and filtering. Hence, the received power, $P_\text{bg}$, into the photodetector due to solar radiation can be modelled as \cite{sun_bg}

\begin{equation}
    P_{\text{bg}} = F_{\lambda}A_{\text{rx}}\eta_{\text{rx}}\Delta \lambda\cdot \frac{\Omega_r}{\Omega_s},
\end{equation}
where the related parameters are described as
\begin{enumerate}
    \item [-] $F_{\lambda}$ is the solar spectral irradiance, typically in [W$\text{m}^{-2}$$\text{nm}^{-1}$].
    \item [-] $\eta_{\text{rx}}$ is the receiver optical efficiency.
    \item [-] $\Delta\lambda$ is the optical bandpass.
    \item [-] $\Omega_s$ is the solid angle of the sun as viewed from LEO.
    \item [-] $\Omega_r = \min{(\Omega_s, \Omega_{\text{FOV}})}$ is the smaller of the receiver FOV solid angle, $\Omega_{\text{FOV}}$, and $\Omega_s$.
\end{enumerate}
Furthermore, the Earth can be modelled as a Lambertian reflector with a $2\pi$ steradian FOV \cite{earth_bg}. Therefore, the received power due to the Earth's albedo is obtained as

\begin{equation}
    P_{\text{bg}}= E_a F_{\lambda}A_{\text{rx}}\eta_{\text{rx}}\Delta \lambda\cdot \frac{\Omega_{\text{FOV}}}{2\pi},
\end{equation}
where $E_a$ is the Earth's albedo.

An APD with excess noise factor $F$, illuminated by signal power $P_{\text{sig}}$ and background power $P_{\text{bg}}$, contributes a shot-noise PSD at the TIA input \cite{Ghassemlooy2019-kk} as
\begin{equation}
\label{eq:17}
N_{\text{sh}}(x,g) = 2qM^2R\bigl(P_{\text{sig}}(x,g) + P_{\text{bg}}\bigr)F.
\end{equation}
In addition, RIN contributes as
\begin{equation}
\label{eq:18}
N_{\text{RIN}}(x,g) = \eta_{\text{RIN}} M^2 R^2 P_{\text{sig}}^2(x,g).
\end{equation}
For noise bandwidth $B$, the received photocurrent $i_{\text{rx}}$ is Gaussian with mean given by \eqref{eq:15} and variance combining illumination-independent and dependent terms
\begin{equation}
\label{eq:19}
\sigma^2_{\text{N}}(x,g) = \bigl(N_{\text{floor}} + N_{\text{sh}}(x,g) + N_{\text{RIN}}(x,g)\bigr)B,
\end{equation}
\begin{equation}
\label{eq:20}
i_{\text{rx}}(x,g) \sim \mathcal{N}\left(\langle i_{\text{rx}}(x,g)\rangle,\sigma^2_{\text{N}}(x,g)\right).
\end{equation}
Since $g$ is a realization of the stochastic channel gain $G$, the conditional distribution of $i_{\text{rx}}$ given the transmitted symbol $x$ is a Gaussian mixture as
\begin{equation}
\label{eq:21a}
\langle i_{\text{rx}}|x \rangle = \mathbb{E}_G[\langle i_{\text{rx}}(x,G)\rangle] = 
\int_0^{\infty}\langle i_{\text{rx}}(x,g)\rangle f_G(g)dg.
\end{equation}
\begin{equation}
\label{eq:21}
p(i_{\text{rx}}|x) =
\int_0^{\infty}\mathcal{N}\left(i_{\text{rx}};\langle i_{\text{rx}}(x,g)\rangle,\sigma^2_{\text{N}}(x,g)\right) f_G(g)dg.
\end{equation}
Here, $\mathcal{N}\left(z;\mu,\sigma^2\right)$ denotes the Gaussian PDF of variable $z$ with mean $\mu$ and variance $\sigma^2$.

\begin{figure}[h]
\centering
\includegraphics[width=0.9\columnwidth]{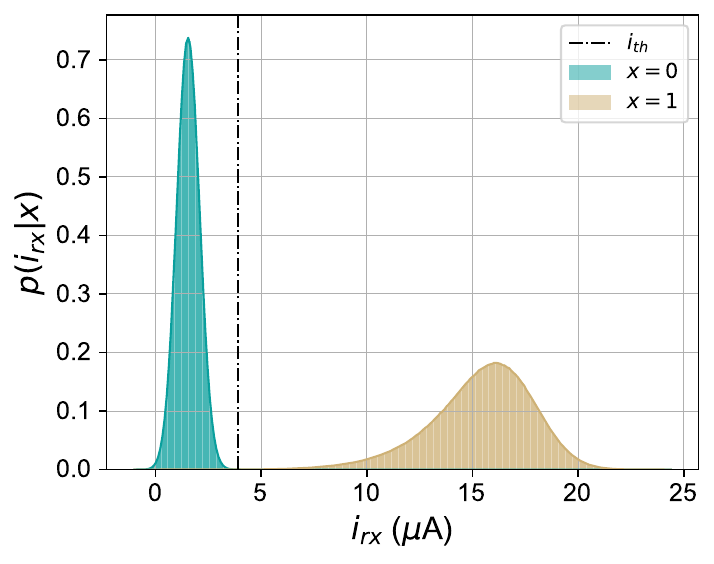}
\caption{Conditional transition distributions $p(i_{\text{rx}}|x)$ for OOK in the MRR channel (400 km LOS), for parameters in table~\ref{table222}. The off-state distribution is near-Gaussian, while the on-state exhibits skew and a long tail due to signal-dependent noise. The ML decision threshold $i_{\text{th}}$ is set at the overlap point.}
\label{fig:i_rx}
\end{figure}

Accordingly, Fig.~\ref{fig:i_rx} shows the numerically obtained transition probability densities for a 400 km LOS link. The off-state current $(x=0)$ is approximately Gaussian, peaking near zero due to modulator insertion loss. The on-state current $(x=1)$ is strongly non-Gaussian, retaining a pronounced left tail from the fading-induced channel gain distribution $f_G(g)$ while while an extended, near-Gaussian right tail is introduced by receiver noise. The retention of the skew in the on-state distribution indicates that fading remains the dominant impairment relative to additive noise, implying a higher signal-to-noise ratio than in the off-state. At the same time, the broadened right tail, approximately Gaussian in shape, indicates a signal-dependent noise variance. These features are typical of a shot-noise-dominated, signal-dependent noise regime. The maximum-likelihood (ML) decision threshold $i_{\text{th}}$, marked in the figure, corresponds to the point of overlap between the two distributions, which determines the minimum achievable BER. 

\subsection{Performance Metrics}\label{sec:metrics}

For equiprobable symbols, the optimal ML decision threshold minimizes the ensemble average BER. It can be mathematically defined as the point of intersection of the two conditional PDFs as
\begin{equation}
\label{eq:22}
    p\left(i_{th}|0\right) = p\left(i_{th}|1\right),
\end{equation}
Thus, the corresponding optimal uncoded BER can be written as
\begin{equation}
\label{eq:23}
    \bar{P}_b = \frac{1}{2}\left[1-F_{i_{\text{rx}}|X}\left(i_{th}|0\right) + F_{i_{\text{rx}}|X}\left(i_{th}|1\right)\right].
\end{equation}
For performance evaluation, we define the outage probability in terms of the information-theoretic capacity of a binary input, continuous output memoryless channel. In conventional RF or optical coherent systems, outage is typically characterized by an SNR threshold, below which reliable communication cannot be sustained. However, in IM-DD systems employing OOK modulation, the noise variance is signal dependent. Specifically, both shot noise and relative intensity noise scale with the instantaneous received optical power. As a result, a single well‑defined SNR cannot be assigned to the channel, and standard SNR‑based outage definitions may therefore fail to capture the system performance. Instead, channel capacity provides a rigorous measure of the maximum achievable rate under these noise statistics. The outage probability is therefore defined as the probability that the instantaneous channel capacity falls below a target transmission rate.

The instantaneous channel capacity is defined as the maximum mutual information, maximized over all possible input distributions, that is,
\begin{equation}
\label{e:24}
C(g) = \max_{P_X(0)\in [0,1]}\left[h(i_{\text{rx}}|G=g) - h(i_{\text{rx}}|X,G=g)\right],
\end{equation}
where $h(\cdot)$ denotes the differential entropy, given by
\begin{equation}
\label{eq:25}
h(i_{\text{rx}}|G=g) = -\int_{-\infty}^{\infty} p(i_{\text{rx}}|G=g)\log p(i_{\text{rx}}|G=g) di_{\text{rx}}.
\end{equation}
The conditional distribution of the received current is
\begin{equation} \label{eq:26} p(i_{\text{rx}}|G=g) = \sum_{x\in \{0,1\}}P_X(x)\mathcal{N}(i;\langle i_{\text{rx}}(x,g)\rangle,\sigma^2_N(x,g)), \end{equation}
and the conditional entropy can thus be obtained as
\begin{equation} \label{eq:27} h(i_{\text{rx}}|X,G=g) = \sum_{x\in \{0,1\}}P_X(x)\frac{1}{2}\log{\left(2\pi e\sigma^2_N(x,g)\right)}. \end{equation}
Accordingly, the outage probability is defined as 
\begin{equation} \label{eq:28} P_{\mathrm{out}} = \mathbb{P}_G\left[C(G) < C_{\mathrm{out}}\right], \end{equation}
where $C_{\text{out}}$ denotes the target outage capacity.

Moreover, in fading channels, system performance is commonly characterized using a combination of BER and outage probability. With strong forward error correction (FEC), error free communication, corresponding to an average post FEC BER on the order of $10^{-15}$, can typically be achieved provided that the uncoded BER remains below a scheme dependent threshold \cite{AIR}. When the interleaving depth is sufficiently large to ensure uncorrelated errors, the BER metric alone is adequate. However, when the interleaving depth is shorter than the channel coherence time, error bursts may occur, and outage probability provides a complementary measure of communication reliability.

A limitation of this approach is its dependence on arbitrary design choices, including the specific FEC scheme and the selected outage threshold. Channel performance may therefore appear artificially favorable under certain assumptions. To address this issue, the authors in \cite{AIR} proposed the achievable information rate as a modulation and code independent performance metric. In this work, we adopt the AIR as the primary optimization metric, as it captures the effect of channel statistics in a single rigorous quantity without the need for threshold based definitions.

Under the assumption of sufficiently deep interleaving, the optical channel described in \eqref{eq:21} can be modeled as a binary input, continuous output memoryless channel. For equiprobable inputs and maximum likelihood decision making, the AIR reduces to the mutual information between the transmitted symbol $X$ and the received photocurrent $i_{\text{rx}}$,
\begin{equation} \label{e:29} I(X;i_{\text{rx}}) = h(i_{\text{rx}}) - h(i_{\text{rx}}|X), 
\end{equation}
where the entropy terms are given by
\begin{equation} \label{eq:30} h(i_{\text{rx}}) = -\int_{-\infty}^{\infty}p(i_{\text{rx}})\log{p(i_{\text{rx}})}di_{\text{rx}}, \end{equation}
\begin{equation}
\label{eq:31} p(i_{\text{rx}}) = \frac{1}{2}\sum_{x\in \{0,1\}}p(i_{\text{rx}}|x), 
\end{equation}
and
\begin{equation} \label{eq:32} h(i_{\text{rx}}|X) = -\frac{1}{2}\sum_{x\in \{0,1\}}\Biggl[\int_{-\infty}^{\infty}p(i_{\text{rx}}|x)\log{p(i_{\text{rx}}|x)}di_{\text{rx}}\Biggr]. \end{equation}
\section{Simulation-Based Channel Optimisation}\label{sec:results}

As thoroughly discussed, the retroreflector-based optical link involves multiple interdependent parameters whose combined influence on system performance cannot be captured by closed-form analysis. Because the channel gain distribution and noise statistics lack tractable expressions, we employ Monte Carlo simulations to evaluate link performance and guide design optimisation under CubeSat SWaP constraints. The simulation parameters are summarized in Table \ref{tab:1}.

\begin{table}[t]
\centering
\caption{Core Link Parameters Used in Optimisation}
\label{tab:1}
\begin{tabular}{|l|l|l|}
\hline
\textbf{Name} & \textbf{Parameter} & \textbf{Value} \\
\hline\hline
Number of trials & $N$ & $5\times10^5$ \\
Transmit power & $P_t$ & 2 W \\
Optical Wavelength & $\lambda$ & 850 nm \\
Modulation Bandwidth & $B$ & 1 GHz \\
Retroreflector Diameter & $D_{\text{rr}}$ & 10 cm \\
Divergence (1/$e^2$ full angle) & $\theta_{\text{div}}$ & 10 $\mu$rad \\
Pointing error (Tx/Rx) & $\sigma_{tx}, \sigma_{rx}$ & 1 $\mu$rad \\
CubeSat pointing error & $3\sigma_{\text{rr}}$ & $1^\circ$ \\
\hline
\end{tabular}
\end{table}

\begin{figure}[h]
\centering
\includegraphics[width=0.9\columnwidth]{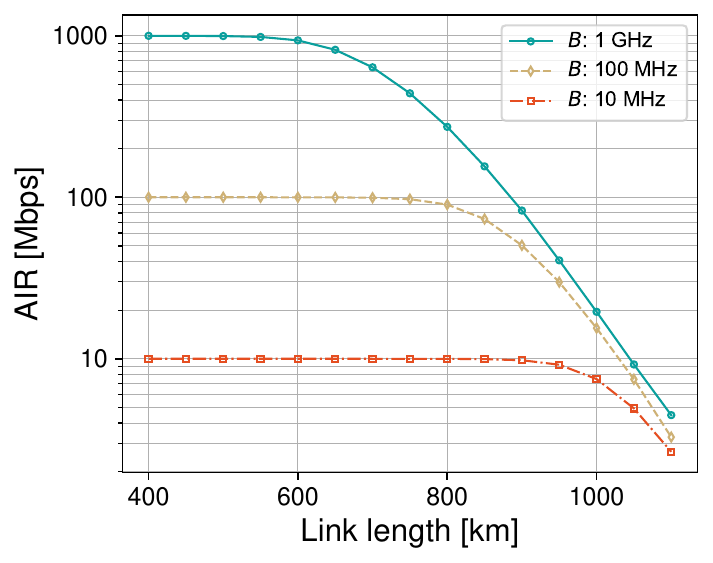}
\caption{AIR (Mbps) versus link length for OOK MRR links, across matched modulator and receiver bandwidths $B$.}
\label{fig:BW}
\end{figure}

Fig.~\ref{fig:BW} illustrates the effect of modulation bandwidth on AIR as a function of link distance. As expected, higher bandwidth enables higher symbol rates, but also increases noise power in proportion to $B$. This creates an inverse trade-off, i.e., reducing bandwidth lowers noise variance, decreases distributional overlap \eqref{eq:21}, and thereby improves AIR at longer ranges. With practical MRR-based EAMs supporting GHz operation \cite{MQW_2}, near-error-free communication is feasible up to approximately 600 km at $B=1$ GHz.

\subsection{Tx/Rx Optimisation}

The performance of the retroreflector-based link depends critically on the transmitter and receiver design choices. In particular, transmit power, beam divergence, and receiver FOV determine the achievable range and reliability. Perfromance evaluation based on these parameters are presented below.

\subsubsection{Transmit Power}
Fig.~\ref{fig:P_t} shows the AIR as a function of link length for different transmit powers. At low powers, the maximum operational range scales approximately as $P_t^{1/4}$, consistent with quartic channel gain decay; for example, doubling $P_t$ from 1 W to 2 W extends the range by about 19\%. At higher powers, however, this scaling saturates because noise variances from shot noise and RIN also grow with the received signal. Once the transmit power exceeds a few watts, further increases yield only marginal improvements, indicating that power scaling is an inefficient strategy for CubeSat‑class platforms.

\begin{figure}[h]
\centering
\includegraphics[width=0.9\columnwidth]{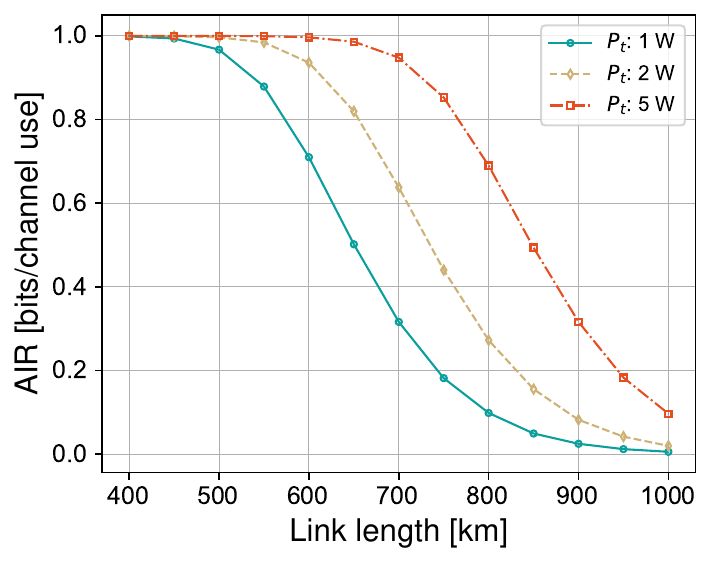}
\caption{AIR (bits/channel use) versus LOS link length for different interrogator transmit powers $P_t$. At higher $P_t$, returns diminish due to signal-dependent noise.}
\label{fig:P_t}
\end{figure}

\subsubsection{Beam Divergence}
\begin{figure}[h]
\centering
\includegraphics[width=0.9\columnwidth]{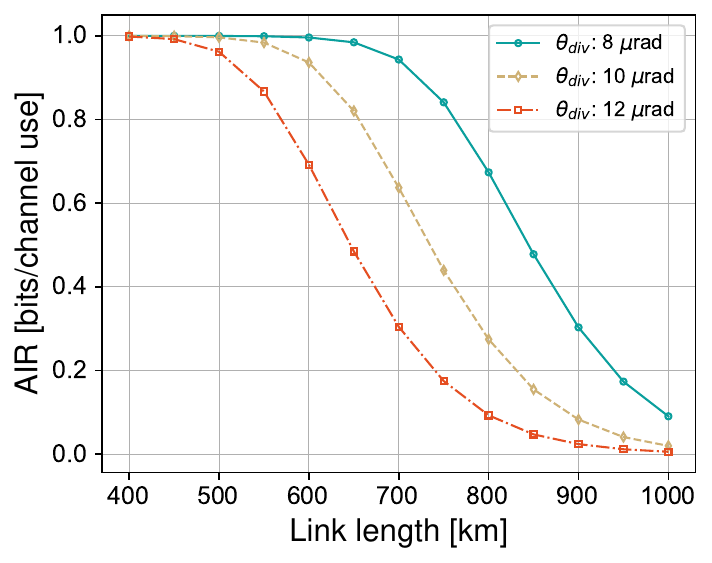}
\caption{AIR versus LOS link length for diffraction-limited divergences $\theta_{\text{div}}$. Narrow beams extend range but heighten pointing sensitivity.}
\label{fig:theta_div}
\end{figure}

The impact of interrogator beam divergence is shown in Fig.~\ref{fig:theta_div}. For a co-boresighted transmit/receive aperture, on-axis gain scales approximately as $\theta_{\text{div}}^{-4}$, yielding longer ranges for narrower beams. However, very small divergences increase susceptibility to pointing errors, which can nullify the gain advantage. Practical systems are also limited by PAT capabilities, which typically constrain divergence to the $10$–$100~\mu$rad range \cite{HEMMATI2013121}. The design trade-off therefore lies in selecting the narrowest divergence compatible with the terminal’s pointing accuracy.

\subsubsection{Background Illumination}
\begin{figure}[h]
\centering
\includegraphics[width=0.9\columnwidth]{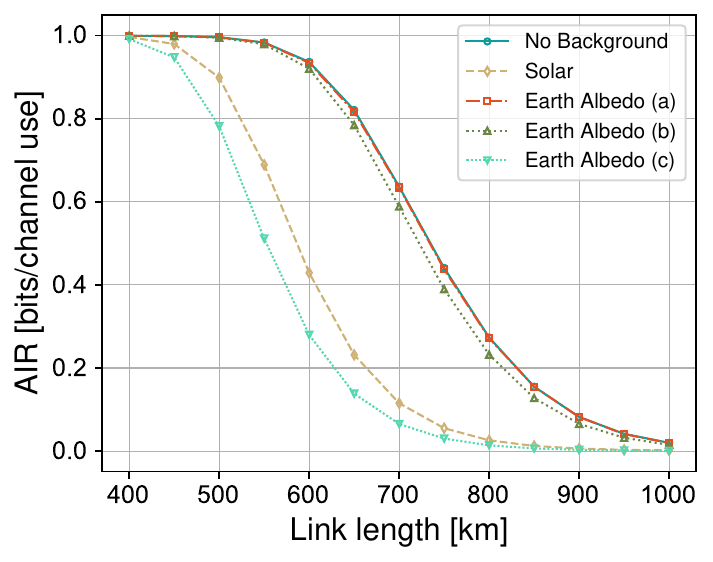}
\caption{\textcolor{black}{AIR versus LOS link length under different background conditions: negligible background, solar irradiance, and Earth albedo. (a) Narrow FOV ($\theta_{\text{FOV}}=100$ $\mu\mathrm{rad}$). (b) Moderate FOV ($\theta_{\text{FOV}}=1$ mrad). (c) Wide FOV ($\theta_{\text{FOV}}=10$ mrad). Narrow FOV reduces background sensitivity.}}
\label{fig:Bg}
\end{figure}

Fig.~\ref{fig:Bg} compares link performance under negligible background interference, direct solar irradiance, and Earth albedo. Solar illumination severely degrades performance even at short ranges, emphasizing the importance of scheduling to avoid Sun-in-FOV conditions \cite{solar_schedule}. \textcolor{black}{Albedo has a weaker impact, but becomes significant for wide FOV receivers. More precisely, for $\theta_{\text{FOV}}=100$ $\mu\mathrm{rad}$, the albedo contribution is negligible, whereas increasing $\theta_{\text{FOV}}$ into the 1-10 mrad range leads to a significant performance reduction. At $\theta_{\text{FOV}} = 1$ mrad, Earth albedo becomes more detrimental than direct solar illumination. Thus, optimal design favors a narrow receiver FOV, balanced against the pointing error tolerance requirements, and on-axis gain (as discussed in Figs.~\ref{fig:g_rx} and \ref{fig:L_pd_2}).}

In a nutshel, the findings suggest that retroreflector‑enabled OISLs achieve optimal performance under moderate transmit power, narrow beam divergence compatible with PAT limits, and a small receiver FOV. These parameters jointly extend the achievable range while maintaining robustness against noise, pointing errors, and ambient illumination.


\subsection{Retroreflector Optimisation}

The performance of retroreflector-based OISLs is shaped by CubeSat pointing accuracy, retroreflector aperture size, orbital geometry, and device-level modulation characteristics. Simulation results in this subsection highlight the trade-offs associated with each factor.

\subsubsection{Pointing Accuracy}
\begin{figure}[h]
\centering
\includegraphics[width=0.9\columnwidth]{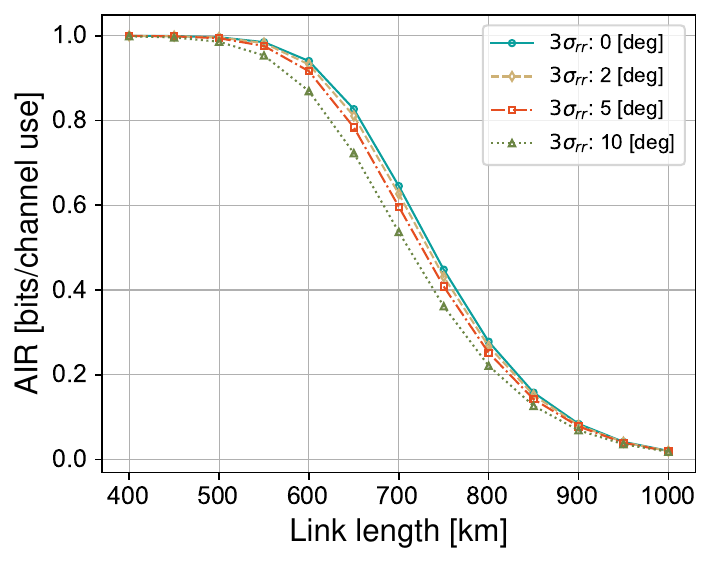}
\caption{AIR versus LOS link length for CubeSat ADCS pointing errors, assuming a wide-field focal-plane modulator. Degree-level errors cause negligible degradation.}
\label{fig:sigma_rr}
\end{figure}

Fig.~\ref{fig:sigma_rr} shows that CubeSat pointing errors have little effect on performance as long as $3\sigma < 2^{\circ}$. Since many CubeSat ADCS systems routinely achieve $3\sigma < 1^{\circ}$, the MRR architecture is naturally compatible with CubeSat platforms. Its resilience to degree-level pointing errors is a key advantage compared with laser transmitters, which require sub-10~$\mu$rad precision.

\subsubsection{Retroreflector Aperture}
\begin{figure}[h]
\centering
\includegraphics[width=0.9\columnwidth]{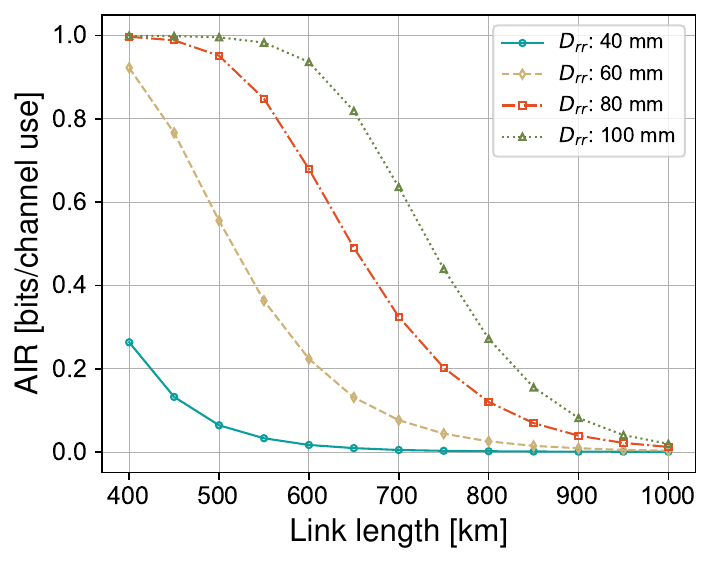}
\caption{AIR versus LOS link length for different retroreflector diameters. Larger apertures extend range but heighten velocity aberration losses.}
\label{fig:d_rr}
\end{figure}

The retroreflected beam follows an Airy profile, with divergences in the 10–100~$\mu$rad range for the apertures considered. Unlike conventional laser terminals, which would demand PAT systems accurate to $\lesssim 10~\mu$rad, the retroreflector decouples beam divergence from PAT requirements. Fig.~\ref{fig:d_rr} illustrates the dependence of AIR on retroreflector diameter. In principle, link length scales linearly with aperture size due to the quartic dependence of gain. However, very large apertures narrow the Airy profile, increasing sensitivity to velocity aberration (Fig.~\ref{fig:L_rr}). For CubeSats, practical limits cap the aperture diameter at 100 mm, which we adopt as the design baseline.

\subsubsection{Orbital Inclination}
\begin{figure}[h]
\centering
\includegraphics[width=0.9\columnwidth]{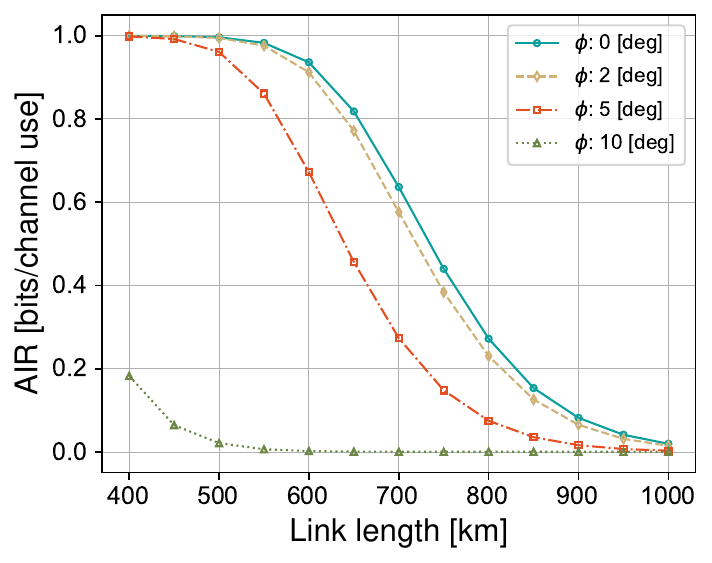}
\caption{AIR versus LOS link length for different orbital angular separations $\phi$. Small angular separations are tolerable; larger separations severely limit performance.}
\label{fig:phi}
\end{figure}

The effect of orbital geometry is shown in Fig.~\ref{fig:phi}. For orbital angular separations $\phi < 2^{\circ}$, performance remains largely unaffected. Beyond this, velocity aberration dominates, limiting feasible link lengths. At $\phi > 5^{\circ}$, performance collapses above 500 km, suggesting that retroreflector-enabled OISLs are best suited to nearly co-planar orbits.

\subsubsection{Wavelength and Modulator Choice}
\begin{figure}[h]
\centering
\includegraphics[width=0.9\columnwidth]{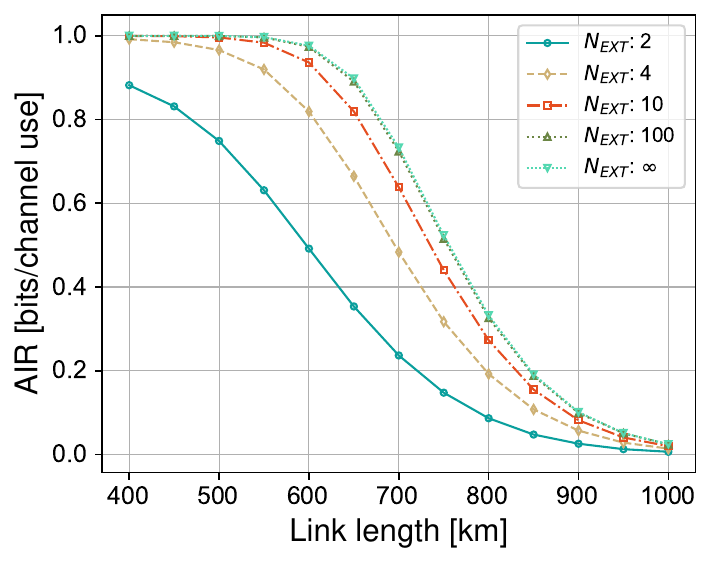}
\caption{AIR versus LOS link length for different modulator extinction ratios $N_{\text{EXT}}$. GaAs modulators at 850 nm outperform InP devices at 1550 nm.}
\label{fig:N_EXT}
\end{figure}

Unlike terrestrial or downlink retroreflector systems, where 1550 nm is preferred for eye safety and atmospheric transparency, inter-satellite links are not atmosphere-limited. This enables the use of shorter wavelengths such as 850 nm, which offer narrower diffraction-limited divergences and compatibility with high-performance GaAs-based MQW modulators. Fig.~\ref{fig:N_EXT} demonstrates the impact of modulator extinction ratio $N_{\text{EXT}}$ on AIR. GaAs devices, with contrast ratios up to 20 dB and insertion losses around 6 dB, outperform InP-based devices at 1550 nm, which are limited to $N_{\text{EXT}} \lesssim 2$. An extinction ratio of $N_{\text{EXT}}=10$ provides a practical trade-off between performance and power consumption, while $N_{\text{EXT}}=100$ represents the upper bound on achievable performance.

The results of this section show that MRR‑enabled CubeSat OISLs can tolerate degree‑level pointing errors, operate effectively with apertures up to 100 mm, and achieve long‑range performance when orbital angular separations are small. Moreover, by leveraging GaAs modulators at 850 nm, CubeSats can attain high extinction ratios and narrower beams without relying on complex PAT systems, making retroreflector‑based links a promising solution for SWaP‑constrained inter‑satellite communications.

\section{Resource Budgets}\label{sec:resource}
After completing the performance evaluation, we now quantify the size and power requirements of the proposed MRR‑enabled CubeSat terminal, ensuring compatibility with typical CubeSat SWaP constraints. In addition to complementing the analytical models and simulation results presented earlier, this resource analysis offers practical insight into the feasibility of implementation. It also provides a foundation for comparison with state‑of‑the‑art laser‑based CubeSat platforms, which we revisit in Section \ref{sec:comp}. 
\subsection{Size Budget}\label{sec:weight}
The optical aperture is the dominant factor in determining the size of the MRR-enabled CubeSat terminal. A 10 cm diameter aperture, the maximum practical size for CubeSat designs, paired with f/1.5 optics yields a minimum optical train length of approximately 15 cm. Including mechanical housing and retroreflector assembly, the complete module requires a volume of roughly 2U, leaving sufficient margin for supporting electronics. Within a 6U CubeSat, the module would therefore occupy about one-third of the available volume, ensuring compatibility with typical spacecraft layouts.

\subsection{Power Budget}\label{sec:power}

The primary power consumers in the terminal are the MQW modulator array and the high-speed FPGA driving. These parameters are evaluated in the following.

\subsubsection{Modulator capacitance}
Each MQW pixel can be modeled as a stack of parallel-plate capacitors in series \cite{MQW_Cap} as
\begin{equation}
\label{eq:cap}
\frac{1}{C} = \sum_i \frac{d_i}{\epsilon_i \epsilon_0 A},
\end{equation}
where $d_i$ and $\epsilon_i$ are the thickness and dielectric constant of layer $i$, $\epsilon_0$ is the vacuum permittivity, and $A$ is the pixel area. For GaAs-based devices \cite{MQW_1}, the capacitance density is approximately $100 ~\text{pF/mm}^2$.

\subsubsection{Bandwidth and pixel sizing}
The 1 GHz RC bandwidth limit, assuming a 50~$\Omega$ driver, constrains the maximum pixel capacitance to 3 pF, corresponding to a maximum pixel area of $0.03 \text{mm}^2$. To achieve a $1^{\circ}$ field of view with a 10 cm, f/1.5 aperture, the modulator array must have a diameter of at least 5 mm, or $21 \text{mm}^2$ total area \cite{Quintana2017-gc}. With a 70\% fill factor, this corresponds to approximately 500 pixels.

\subsubsection{Dynamic power}
The dynamic switching power of a single pixel is
\begin{equation*}
P = \tfrac{1}{2}CV^2f_c,
\end{equation*}
which yields 1.5 mW per pixel for $C=3$ pF, $V=1$ V, and $f_c=1$ GHz. For 500 pixels, the array consumes about 0.75 W of useful switching power.

\subsubsection{Driver efficiency}
The upper bound on driver consumption can be estimated by modeling the pixel load as a matched resistor $R$ driven by a sinusoidal waveform
\begin{equation*}
P = \frac{V^2}{8R}.
\end{equation*}
For $R=50~\Omega$ and $V=1$ V, this gives 2.5 mW per pixel, implying a driver efficiency of ~60\%. Thus, the complete 500-pixel array consumes approximately 1.25 W.

\subsubsection{FPGA consumption}
The high-speed FPGA responsible for signal generation represents the second major power draw. Based on similar CubeSat optical payloads \cite{Kingsbury2015}, we allocate ~1.25 W for FPGA and digital electronics.

\subsubsection{Total power budget}
The overall terminal power consumption is therefore estimated at 2.5 W, well within the typical 5–10 W payload allocations available on modern 6U CubeSats.

\section{System Comparison}\label{sec:comp}

To contextualize the proposed retroreflector-enabled inter-satellite link, we compare its performance against three representative CubeSat OISL systems: CLICK B/C, OCSD, and OSIRIS4CubeSat. For fairness, all links are modeled with an identical receiver terminal. In the following, we briefly review the specifications of each benchmark.

\subsubsection{CLICK B/C}
Scheduled for launch in 2026, CLICK B/C represents a high-performance CubeSat crosslink terminal \cite{CLICKB_C}. It employs a two-stage pointing system: coarse body-pointing with $0.15^{\circ}$ ADCS accuracy ($3\sigma$), followed by beacon-aided fine pointing using a MEMS fast steering mirror with demonstrated $16  \mu$rad precision ($1\sigma$). The system uses a 71~$\mu$rad FWHM divergence Gaussian beam with 200 mW transmit power and consumes 30 W within a $<2$U form factor \cite{CLICK_A, CLICK_2}.

\subsubsection{OCSD}
Launched in 2017, it showcased a body-pointing-only laser downlink terminal \cite{OCSD}. With ADCS performance of $0.024^{\circ}$ ($3\sigma$), the system relied on a relatively wide $0.06^{\circ}$ FWHM divergence beam and a 2 W source. This reduced PAT complexity but imposed a higher pointing accuracy requirement. The terminal consumed 10–20 W and occupied a 1.5U CubeSat.

\subsubsection{OSIRIS4CubeSat}
The OSIRIS4CubeSat terminal is a compact, low-SWaP commercial optical downlink solution \cite{OSIRIS4CubeSat}. It employs beacon-aided pointing and transmits via a 200~$\mu$rad full-angle Gaussian beam with 100 mW output power. While detailed pointing error figures are unavailable, we assume $50 \mu$rad ($3\sigma$) based on beamwidth. The terminal consumes 8.5 W and occupies 0.3U, making it one of the most size-efficient optical terminals.

\subsection{Comparative Results}
Performance comparisons are presented in Fig.~\ref{fig:comparison}. The retroreflector-based system is evaluated at two operating points: (a) a 2 W interrogating laser, and (b) a 5 W source. Simulation parameters are also provided in Table IV.
\subsubsection{AIR}
Figs.~\ref{fig:comparison}(a)–(b) reveal a fundamental scaling difference. Conventional one-way OISLs (CLICK, OCSD, OSIRIS) exhibit inverse-square dependence on link length, while the retroreflector-based round-trip link decays with the fourth power of distance (as described in \eqref{eq:g2}). Consequently, the retroreflector system degrades more rapidly with distance. Nonetheless, at short ranges ($z \lesssim 500$ km), its performance is comparable to OSIRIS and superior to OCSD. CLICK remains the strongest performer due to its narrow beams and fine-pointing PAT.
\subsubsection{BER and outage probability}
\begin{figure*}[t]
\begin{center}
\includegraphics[width=\textwidth]{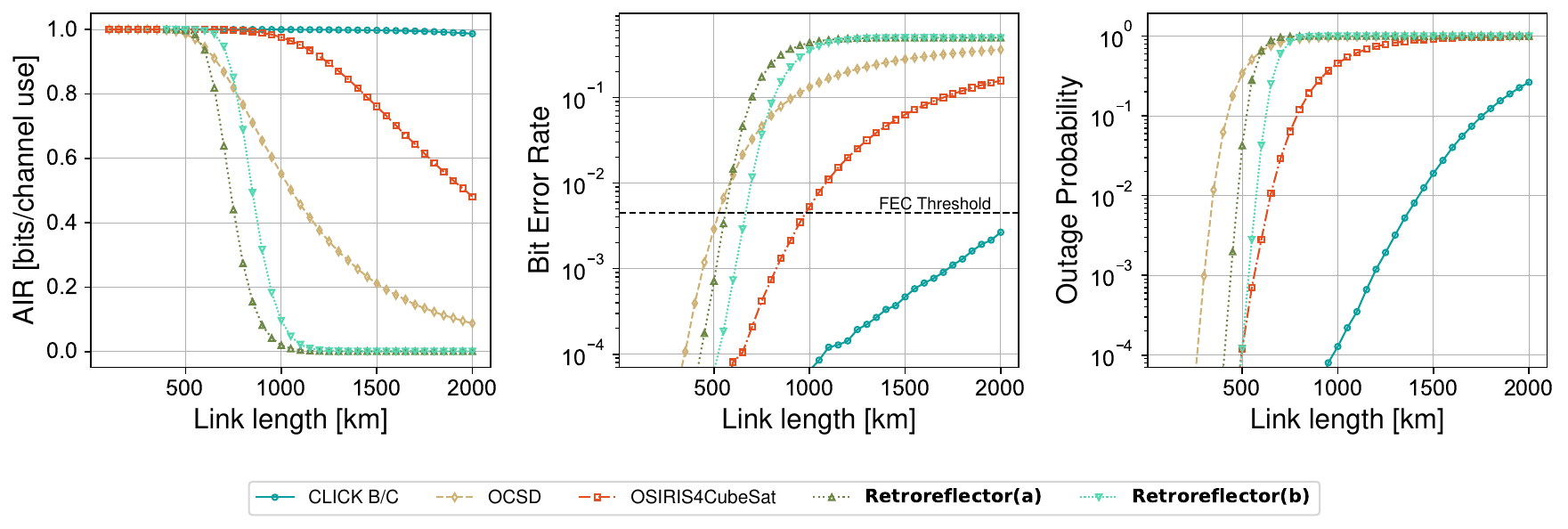}
\end{center}
\caption{Performance comparison of the proposed retroreflector link with CLICK B/C, OCSD, and OSIRIS4CubeSat transmit terminals, using the same receiver terminal (Table \ref{table222}) for consistent evaluation. (a) AIR versus link length. (b) BER versus link length, with FEC threshold $4.5\times10^{-3}$ \cite{ITU-T-G975.1-2004-secI.5.3}. (c) Outage probability versus link length, with threshold defined by the channel capacity at the FEC bit-error-rate threshold.}
\label{fig:comparison}
\end{figure*}

Using an FEC threshold of $4.5\times10^{-3}$ \cite{ITU-T-G975.1-2004-secI.5.3}, the retroreflector-based link achieves maximum ranges of 600–700 km, on par with OCSD and slightly below OSIRIS. Outage probability results in Fig.~\ref{fig:comparison}(c) confirm this trend, i.e., at ranges of 400–500 km, retroreflector-based links achieve $\mathbb{P}_{\text{out}} < 10^{-4}$, rendering them competitive with OSIRIS under realistic outage conditions.

Indeed, CLICK’s superior performance reflects its design as a symmetric crosslink terminal equipped with high‑power PAT, making it unsuitable as a baseline for SWaP‑limited CubeSats. In contrast, the retroreflector‑based architecture aligns more naturally with asymmetric systems such as OSIRIS and OCSD, which rely on simplified terminals and ground‑ or mothership‑based asymmetry. Within this design space, the retroreflector approach offers comparable range and reliability while eliminating onboard laser transmitters and PAT subsystems, thereby significantly reducing SWaP demands.

\section{Conclusion and Future Work}\label{sec:conclusion}
This paper presented a statistical channel model and performance analysis of an OOK-modulated, MRR-enabled asymmetric CubeSat inter-satellite link. By incorporating stochastic pointing errors, velocity aberration, and signal-dependent noise, we derived BER, outage probability, and AIR metrics, and used them to optimize performance under CubeSat SWaP constraints. Results show that GHz-class GaAs MQW modulators can support reliable CubeSat OISLs at ranges up to 500–700 km, outperforming OCSD and comparable to OSIRIS4CubeSat, while requiring only ~2.5 W and a compact 2U form factor. The approach is uniquely resilient to CubeSat-level pointing errors, offering a practical low-SWaP alternative to conventional laser terminals. However, the round-trip geometry amplifies range dependence, with velocity aberration and aperture size emerging as key limiting factors. It was shown that retroreflector-enabled OISLs are not a replacement for high-performance crosslink systems like CLICK, but provide a complementary, low-power solution for short-range asymmetric CubeSat links. 

Future research should investigate adaptive beam shaping or optical pre‑compensation techniques to mitigate velocity aberration, hybrid RF–optical architectures to extend operational range, and tailored FEC schemes that leverage the unique channel statistics of MRR‑based links. Experimental validation through hardware prototypes and in‑orbit demonstrations will be essential to assess the practicality of these systems and to inform their integration into future CubeSat missions.
\section*{APPENDIX}
\label{sec:app}

\begin{table}[ht]
\centering
\caption{Link Parameters Used in Simulation}
\label{table222}
\begin{tabular}{|l|l|l|}
\hline
\textbf{Name} & \textbf{Parameter} & \textbf{Value} \\
\hline\hline
\multicolumn{3}{|l|}{\textit{Simulation Setup}} \\
\hline
Number of trials & $N$ & $5\times10^5$ \\
Bit error rate FEC threshold & $BER_{\text{th}}$ & $4.5\times10^{-3}$ \cite{ITU-T-G975.1-2004-secI.5.3}\\
\hline
\multicolumn{3}{|l|}{\textit{Geometry}} \\
\hline
CubeSat Altitude & $h_1$ & 400 km \\
Interrogator Altitude & $h_2$ & 800 km \\
orbital angular separation & $\phi$ & $0^\circ$ \\
\hline
\multicolumn{3}{|l|}{\textit{Transmitter (Interrogator)}} \\
\hline
Optical Wavelength & $\lambda$ & 850 nm \\
Transmit power & $P_t$ & 2 W \\
Divergence (1/$e^2$ full angle) & $\theta_{\text{div}}$ & 10 $\mu$rad \\
Beamwaist-to-aperture ratio & $\alpha$ & 1.12\\
Pointing error & $\sigma_{tx}, \sigma_{rx}$ & 1 $\mu$rad \\
System optical efficiency & $\eta_{\text{opt}}$ & 3 dB \\
\hline
\multicolumn{3}{|l|}{\textit{Retroreflector Terminal (CubeSat)}} \\
\hline
Retroreflector Diameter & $D_{\text{rr}}$ & 10 cm \\
Pointing error (post-ADCS) & $3\sigma_{\text{rr}}$ & $1^\circ$ \\
\hline
\multicolumn{3}{|l|}{\textit{Receiver / Detector}} \\
\hline
Photodetector & -- & Thorlabs APD430A \\
Excess Noise Factor & $F$ & 4 \\
Receiver field of view & $\theta_{\text{FOV}}$ & 100 $\mu$rad \\
Optical Bandpass & $\Delta\lambda$ & 1 nm \\
\hline
\multicolumn{3}{|l|}{\textit{Modulator}} \\
\hline
Modulation Bandwidth & $B$ & 1 GHz \\
Insertion loss & $\eta_{\text{mod}}$ & 6 dB \\
Extinction ratio & $N_{\text{EXT}}$ & 10 \\
\hline
\multicolumn{3}{|l|}{\textit{Environment}} \\
\hline
Solar spectral irradiance & $F_\lambda$ & 0.96 Wm$^{-2}$nm$^{-1}$ \\
Solar solid angle & $\Omega_{\text{sol}}$ & $6.8\times10^{-5}$ sr \\
Earth albedo & $E_a$ & 0.3 \\
\hline
\end{tabular}
\end{table}

\subsection*{Relative Velocity Derivation}

\begin{figure}[h]
\begin{center}
\includegraphics[width=0.9\columnwidth]{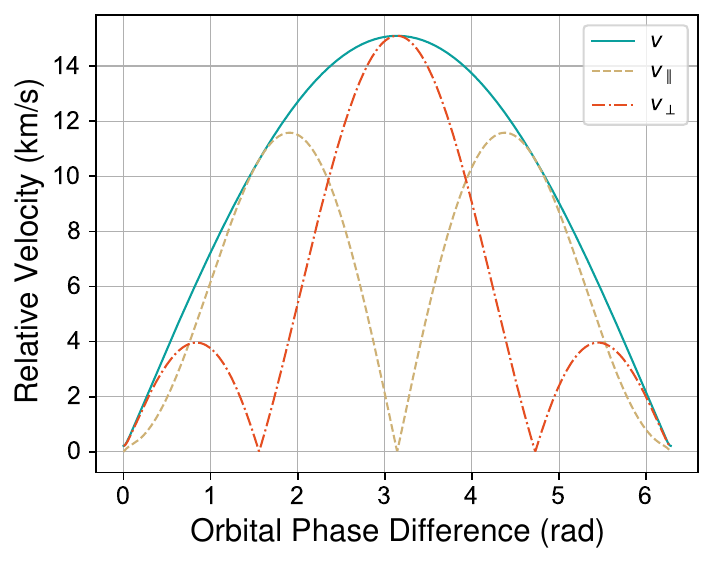}
\end{center}
\caption{Relative velocity $v$ versus along-track phase difference $\theta$ for two satellites with orbital radii of 400 km and 800 km. Components parallel ($v_{\parallel}$) and perpendicular ($v_{\perp}$) to the line of sight are also shown.}
\label{fig:relative velocity}
\end{figure}

We consider two satellites in circular Earth orbits: Satellite A at orbital radius $R_A$ and Satellite B at $R_B$, corresponding to an altitude separation $\Delta h$. Their orbital angular velocities follow from Kepler’s third law as follows
\begin{equation}
    \omega_A = \sqrt{\frac{\mu_E}{R_A^3}}, \hspace{1cm} \omega_B = \sqrt{\frac{\mu_E}{R_B^3}}.
\end{equation}
Here, $\mu_E$ is Earth's standard gravitational parameter.
The relative geometry is described by two parameters, i.e.,
\begin{enumerate}
    \item [-] $\phi$: the angular separation between each orbital plane
    \item [-] $\theta$: the along-track phase difference in their relative orbital planes
\end{enumerate}
Without loss of generality, the initial position vectors can be written as
\begin{equation*}
    \mathbf{r}_A = R_A\begin{pmatrix}
        1\\
        0\\
        0
    \end{pmatrix}, \hspace{1cm}
    \mathbf{r}_B = R_B\begin{pmatrix}
        \cos{\theta}\\
        \cos{\phi}\sin{\theta}\\
        \sin{\phi}\sin{\theta}
    \end{pmatrix},
\end{equation*}
The relative position of B with respect to A is then
\begin{equation*}
        \mathbf{r}_{B/A} = \begin{pmatrix}
        R_B\cos{\theta}-R_A\\
        R_B\cos{\phi}\sin{\theta}\\
        R_B\sin{\phi}\sin{\theta}
    \end{pmatrix}.
\end{equation*}
Similarly, their initial velocity vectors may be inferred from the position vectors and angular velocities
\begin{equation*}
    \mathbf{v}_A =  R_A\omega_A\begin{pmatrix}
        0\\
        1\\
        0
    \end{pmatrix}, \hspace{1cm}
    \mathbf{v}_B = R_B\omega_B\begin{pmatrix}
        -\sin{\theta}\\
        \cos{\phi}\cos{\theta}\\
        \sin{\phi}\cos{\theta}
    \end{pmatrix}.
\end{equation*}
Thus, the relative velocity is
\begin{equation*}
\mathbf{v}_{B/A} = \begin{pmatrix}
        -R_B\omega_B\sin{\theta}\\
        R_B\omega_B\cos{\phi}\cos{\theta}-R_A\omega_A\\
        R_B\omega_B\sin{\phi}\cos{\theta}
    \end{pmatrix}.
\end{equation*}
The relative velocity may be decomposed into two orthogonal components: that in the direction of LOS, $v_{\parallel}$, contributing to Doppler shifts; and that perpendicular to the LOS, $v_{\perp}$, contributing to velocity aberration.
\begin{equation*}
    v_{\parallel} = \frac{\left|\mathbf{v}_{B/A}\mathbf{\cdot}\mathbf{r}_{B/A}\right|}{\left|\mathbf{r}_{B/A}\right|}, \hspace{1cm}
    v_{\perp} = \frac{\left|\mathbf{v}_{B/A}\mathbf{\times}\mathbf{r}_{B/A}\right|}{\left|\mathbf{r}_{B/A}\right|}.
\end{equation*}
Fig. \ref{fig:relative velocity} illustrates these components versus the phase difference $\theta$ for $R_A=400$ km, $R_B=800$ km, and $\phi=0^\circ$. Both $v_{\parallel}$ and $v_{\perp}$ exhibit periodic dependence on $\theta$, with minima occurring when the velocity and LOS vectors align ($\theta = \pm\pi/2$ for $v_{\perp}$, and multiples of $\pi$ for $v_{\parallel}$).

\begin{figure}[h]
\begin{center}
\includegraphics[width=0.9\columnwidth]{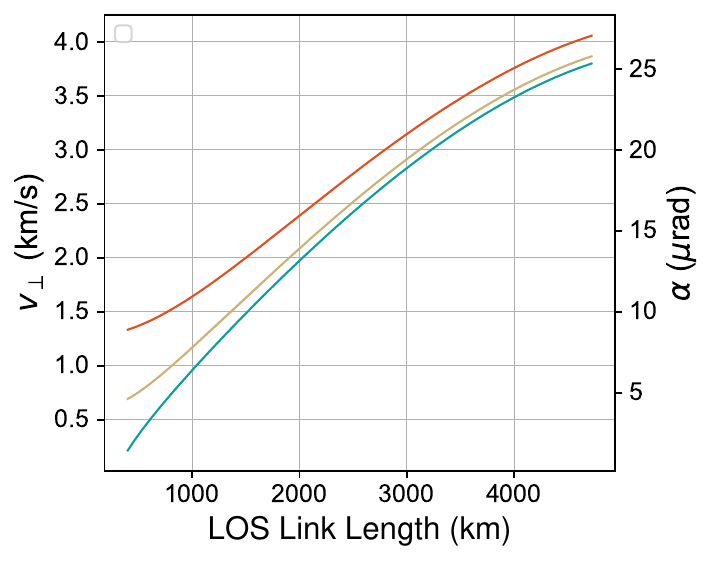}
\end{center}
\caption{\textcolor{black}{Perpendicular relative velocity $v_{\perp}$ (and round-trip angular offset $\alpha$) versus LOS distance for satellites separated by 400 km in orbital radius, shown for various orbital angular separations $\phi$.}}
\label{fig:relative perp velocity}
\end{figure}

The maximum LOS distance is set by Earth blockage. Specifically, the maximum allowable angular separation $\theta_{\text{LOS}}$ is
\begin{equation*}
\theta_{\text{LOS}} = \arccos{\left(\tfrac{R_E}{R_A}\right)} + \arccos{\left(\tfrac{R_E}{R_B}\right)},
\end{equation*}
where $R_E$ is Earth’s radius. For $R_A=400$ km and $R_B=800$ km, this yields a maximum LOS length of $\approx 4750$ km. Fig. \ref{fig:relative perp velocity} plots $v_{\perp}$ as a function of LOS distance for different orbital angular separations $\phi$. Both link length and relative inclination increase $v_{\perp}$, emphasizing their role in velocity aberration and the resulting pointing losses in retroreflector-based OISLs.

\printbibliography

\end{document}